\documentclass[12pt]{article}
\usepackage{amsmath,amsfonts, amssymb,graphicx,geometry,authblk,color} 
\usepackage[ruled]{algorithm2e}

\DeclareMathOperator*{\argmin}{arg\,min}

\newlength\myindent
\title{Accelerated computational micromechanics}
\author{Hao Zhou \& Kaushik Bhattacharya}
\affil{ Division of Engineering and Applied Science\\ California Institute of Technology\\ Pasadena CA 91125}
\date{}

\begin{document}

\maketitle

\begin{abstract}

We present an approach to solving problems in micromechanics that is amenable to massively parallel calculations through the use of graphical processing units and other accelerators.  The problems lead to nonlinear differential equations that are typically second order in space and first order in time.  This combination of nonlinearity and nonlocality makes such problems  difficult to solve in parallel. However, this combination is a result of collapsing  nonlocal, but linear and universal physical laws (kinematic compatibility, balance laws), and nonlinear but local constitutive relations.  We propose an operator-splitting scheme inspired by this structure. The governing equations are formulated as (incremental) variational problems, the differential constraints like compatibility are introduced using an augmented Lagrangian, and the resulting incremental variational principle is solved by the alternating direction method of multipliers.  The resulting algorithm has a natural connection to  physical principles, and also enables massively parallel implementation on structured grids.  We present this method and use it to study two examples. The first concerns the long wavelength instability of finite elasticity, and allows us to verify the approach against previous numerical simulations.  We also use this example to study convergence and parallel performance.  The second example concerns microstructure evolution in liquid crystal elastomers and provides new insights into some counter-intuitive properties of these materials.  We use this example to validate the model and the approach against experimental observations.
\end{abstract}

\section{Introduction} \label{sec:intro}

The recent decades have seen increasing interest in understanding and designing materials with heterogeneous microstructure.  Consequently, a framework of multiscale modeling has emerged, where detailed studies at one scale inform modeling of the behavior at another scale (e.g. \cite{f_book_09,dr_book_11}).  An important tool in this framework -- often described as the homogenization method -- is detailed numerical simulation of the problem at a fine (microscopic) scale using periodic boundary conditions on a representative unit cell to obtain the constitutive behavior at the coarse (macroscopic) scale.  Further, some researchers have explored concurrent methods where the constitutive relation at each point in the computational domain is computed on the fly with a computation at a smaller scale (e.g. \cite{s14}).  For all its promise, such methods are challenging due to the computational cost.

The computational challenges of computational micromechanics have been addressed by algorithmic advances like domain decomposition, software advances like Message Passing Interface (MPI) and hardware advances like fast interconnects in high performance computational clusters.  However, as the size of high performance computational clusters have grown, a recent trend is to incorporate massively parallel accelerators like GPUs to complement the central processing units (CPUs) of the cluster \cite{kirkhwu,exascale}.  Unlike a CPU which contains tens of individually programmable computing cores that can communicate with each other,  an accelerator contains hundreds of computing cores.  However, communication between processors is limited, and groups of these computing cores are restricted to ``single instruction multiple data'' operation where all computing cores perform the same task with different inputs \cite{kirkhwu}.   The current work seeks to exploit computational platforms that incorporate accelerators in studying problems of computational micromechanics. 

A wide variety of phenomena including finite elasticity, crystal plasticity and phase transformations are described by a combination of mechanical displacements and internal variable which are governed by coupled nonlinear differential equations that are typically second order in space and first order in time.  Finite element methods have been widely used to solve such problems (e.g., \cite{yvonnet,rotersetal,seguradoetal}).   Finite element methods are difficult to implement on accelerators, even though there are recent efforts in this direction \cite{gpufem}.   

An alternate approach  began with the pioneering work of Moulinec and Suquet \cite{moulinec1994fast} that exploits the periodic boundary conditions and fast Fourier transforms (FFT).  By introducing a comparison medium, they rewrite the problem of equilibrium in a heterogeneous linear elastic medium to a Lippmann-Schwinger type equation which they solve iteratively using FFT.  Since then, the approach has been applied to many different systems including elasto-plasticity~\cite{moulinec1998numerical}, thermoelasticity~\cite{anglin2014validation}, elasto-viscoplasticity~\cite{lebensohn2012elasto, lebensohn2016numerical}, dislocation~\cite{berbenni2014numerical, bertin2018fft, berbenni2020fast}, piezoelectric materials~\cite{brenner2010computational, vidyasagar2017predicting}, shape-memory polycrystal~\cite{bhattacharya2005model}, and crack prediction of brittle materials~\cite{li2012damage, chen2019fft, schneider2020fft}.   There have also been a number of approaches to accelerate the convergence of FFT-based methods~\cite{eyre1999fast,michel2000computational,monchiet2012polarization,milton2020}.  We refer the reader to Moulinec and Silva \cite{moulinec2014comparison} and Moulinec, Suquet and Milton \cite{msm_ijnme_18} for a discussion.    Other approaches explore Krylov-subspace solvers \cite{zeman2010accelerating,brisard2010fft,brisard2012combining}, Newton's method with conjugate gradient \cite{gelebart2013non,kabel2014efficient}, and quasi-Newton methods \cite{schneider2019barzilai,shanthraj2015numerically, wicht2020quasi}.   Another approach that uses FFT but avoids the use of comparison media are Galerkin-based FFT methods  \cite{vondvrejc2014fft, zeman2017finite, de2017finite, lucarini2019algorithm, ma2019fft, bergmann2020fft} and their variations \cite{lucarini2019dbfft,wicht2020efficient}. Since these are based on FFT, this aspect of these methods can be implemented on GPUs \cite{bertin2018fft}.  Other researchers have used GPUs for the constitutive update \cite{mihaila2014three,knezevic2014high,eghtesad2018spectral}.  Still, these approaches adapt existing algorithms to GPUs.

In this paper, we propose an approach that takes  advantage of the inherent physics and resulting structure of equations to enable massively parallel computations through accelerators.  The approach is discussed in detail in Section \ref{sec:meth}.  Briefly, the key observation is to notice that the nonlinear partial differential equations come about through a {\it composition} of balance laws (mass, momenta, energy) and material behavior.  The balance laws are nonlocal but {\it universal} and {\it linear} if properly formulated. The material behavior is nonlinear and may involve time derivatives, but are {\it local} spatially. Thus, the core difficulty, the combination of nonlinearity and nonlocality, is a result of the composition.  We overcome this by splitting the operator in a novel manner to enable acceleration. Specifically, we start from a (incremental) variational principle, formulate the kinematic compatibility as a constraint using an augmented Lagrangian, and then use the alternating direction method of multipliers (ADMM).  We show that step in the resulting iterative scheme can be implemented trivially in a massively parallel GPU or other accelerator.   The implementation is discussed in Section \ref{sec:imp}.

Our approach leads to a special case of the augmented Lagrangian approach of Michel, Moulinec and Suquet \cite{michel2000computational}.  This was proposed to study problems with large contrast and the method has been used to study voids \cite{lebenvoid} and phase transitions \cite{rlb_actamat_13}.  Originally proposed in small strains, it has been extended to finite strains \cite{shanthraj2015numerically}.  Our method also has similarities to the Hu-Washizu variational principle \cite{w_book_68} and the split Bregman method \cite{go_siamjis_08}.  We discuss these in detail in Section 2. 


We demonstrate the method using two examples. The first, in Section \ref{sec:bif}, concerns the long wavelength instability of finite elasticity, and allows us to verify the approach against previous numerical simulations.  The example also shows how stability analysis with Bloch waves can easily be incorporated into the approach.  We also use this example to study convergence and parallel performance.  The second example, in Section \ref{sec:lce}, concerns microstructure evolution in liquid crystal elastomers and provides new insights into rather counter-intuitive properties of these materials.  We use this example to validate the model and the approach against experimental observations.  We conclude in Section \ref{sec:conc} with a discussion of potential extensions and open issues.

\section{Method} \label{sec:meth}

\subsection{Formulation} \label{sec:form}

It is common in a number of phenomena in solids including plasticity (e.g. \cite{rice_jmps_71}), phase transitions (e.g. \cite{artemev_jin_actamat_01})  and fracture  (e.g. \cite{{bfm_jelas_08}} ) to describe the state of the solid by a deformation gradient $F$ and a set of internal variables $\lambda$ (phase fraction, plastic activity, director field, fracture field etc.). In the absence of inertia, these are governed by a pair of coupled nonlinear partial differential equations  (e.g. \cite{os_cmame_99,mielke}):
\begin{align}
\nabla \cdot \left( W_F (\nabla u, \lambda, x) \right) = 0, \label{eq:elas} \\
W_\lambda (\nabla u, \lambda,x) +  D_v(\lambda_t,x) = 0 \label{eq:int}
\end{align} 
where $u: \Omega \times [0,T] \to {\mathbb R}^3$ is the deformation, $F: \Omega \times [0,T] \to {\mathbb R}^{3 \times 3}$ is the deformation gradient, $\lambda: \Omega \times [0,T] \to {\mathbb R}^d$ is an internal variable or order parameter, $W: {\mathbb R}^{3\times 3} \times {\mathbb R}^d \times \Omega \to {\mathbb R}$ is the stored (elastic) energy density,  $D: {\mathbb R}^d  \times \Omega \to {\mathbb R}$ is the dissipation potential that governs the evolution of the internal variables, $\Omega \subset {\mathbb R}^3$ is the reference domain assumed to be simply connected, $T$ is the final time of interest  and the subscripts denote partial differentiation.  The first of the two equations describes the mechanical equilibrium, and the second the kinetic relation or configurational equilibrium that governs the evolution of the internal variables.  

The first equation above is a second-order nonlinear elliptic partial differential equation in space with time as a parameter.  The second is typically first order in time with space as a parameter, and also nonlinear.  Further, in rate independent phenomena, the dissipation potential may not be continuously differentiable and the second equation is interpreted as a differential inclusion \cite{mielke}.  Thus, these are difficult to solve.  In particular, they require significant amounts of communication to solve in parallel.

However, these equations arise from the agglomeration of a number of simpler equations:
\begin{align}
&\mbox{Compatibility:} & F = \nabla u \iff \mbox{curl } F = 0, \label{eq:compat}\\
&\mbox{Equilibrium:} & \nabla \cdot S = 0, \label{eq:equil}\\
&\mbox{Stress-Strain Relation:} & S = W_F (F,\lambda,x),  \label{eq:ss}\\
&\mbox{Kinetic relation:} & W_\lambda (F, \lambda,x) + D_v (\lambda_t,x) = 0. \label{eq:kr}
\end{align}
Note that the field equations -- the first two -- are linear and universal (in fact Helmholtz projections), while the constitutive updates -- the last two -- are local (albeit nonlinear).  We want to exploit this in the GPU implementation.

In order to do so, consider an implicit time discretization
\begin{align}
&\mbox{Compatibility:} & F^{n+1} = \nabla u^{n+1} \iff \mbox{curl } F^{n+1} = 0,  \\
&\mbox{Equilibrium:} & \nabla \cdot S^{n+1}  = 0, \\
&\mbox{Stress-Strain Relation:} & S^{n+1}  = W_F (F^{n+1},\lambda^{n+1},x),  \\
&\mbox{Kinetic relation:} & W_\lambda (F^{n+1}, \lambda^{n+1},x) + D_v \left( {\lambda^{n+1} - \lambda^n \over \Delta t},x \right) = 0.
\end{align}
This can be written as the following variational problem (e.g., \cite{os_cmame_99}):
\begin{equation} \label{eq:var}
u^{n+1}, \lambda^{n+1} = \argmin  \int_\Omega \left( W( \nabla u, \lambda, x) + \Delta t 
D\left( {\lambda - \lambda^n \over \Delta t},x \right) \right) dx
\end{equation}
or as the following constrained variational problem
\begin{equation}
F^{n+1}, \lambda^{n+1} =  \argmin_{\mbox{\tiny curl F} = 0}   \int_\Omega \left( W( F, \lambda, x) + \Delta t 
D\left( {\lambda - \lambda^n \over \Delta t},x \right) \right) dx.
\end{equation}

We rewrite this constrained variational problems using the {\it augmented Lagrangian} method or method of multipliers (e.g. \cite{g_book_15}).  
Given $\rho > 0$, we seek to find the saddle point:
\begin{equation} \label{eq:al}
\int_\Omega \left( W( F, \lambda, x) + \Delta t  D\left( {\lambda - \lambda^n \over \Delta t},x \right) + \Lambda \cdot (\nabla u - F) + {\rho \over 2} |\nabla u - F|^2   \right) dx
\end{equation}
for $u$, $F$ and the Lagrange multiplier $\Lambda: \Omega \to {\mathbb R}^{3 \times 3}$.
We solve this problem using the {\it alternating direction method of multipliers (ADMM)} \cite{g_book_15,g_chap_16,glowinski1989augmented} which is an iterative method.  

At the (n+1)$^{th}$ time step, given $F^n, \lambda^n, u^n, \Lambda^n$, set $F_0 = F^n, \lambda_0 = \lambda^n, u_0 = u^n, \Lambda_0 = \Lambda^n$ and iterate over $i$

\begin{itemize}

\item {\it Step 1: Local problem.} Update $F, \lambda$ by solving at each $x$
\begin{align}
W_F (F_{i+1},\lambda_{i+1},x) - \Lambda_i - \rho (\nabla u_i - F_{i+1}) = 0, \label{eq:eq}\\
W_\lambda (F_{i+1},\lambda_{i+1},x) + D_v \left( {\lambda_{i+1} - \lambda^n \over \Delta t} ,x \right) = 0 \label{eq:kin}.
\end{align}
 
\item {\it Step 2: Helmholtz projection.}  Update $u$ by solving the partial differential equation
\begin{equation} \label{eq:helm}
- \Delta u_{i+1} =   \nabla \cdot \left(-  F_{i+1}  + {1 \over \rho} \Lambda_i \right) .
\end{equation}

\item {\it Step 3: Update Lagrange multiplier.}  Update $\Lambda$ as
\begin{equation} \label{eq:update}
\Lambda_{i+1} = \Lambda_i + \rho (\nabla u_{i+1} - F_{i+1}) .
\end{equation}

\item {\it Step 4: Check for convergence.}  Check both primal and dual feasibility:
\begin{equation} \label{eq:check}
    r_p := || \nabla u_{i+1} - F_{i+1} ||_{L^2} \le r_p^\text{tolerance}, \quad r_d := \rho / \mu || \nabla u_{i+1} - \nabla u_{i} ||_{L^2} \le r_d^\text{tolerance}
\end{equation}
for given $r_p^\text{tolerance}, r_d^\text{tolerance}$ and representative elastic modulus $\mu$.

\end{itemize}
until convergence and update $F^{n+1} = F_{i}, \lambda^{n+1} = \lambda_i,  u^{n+1} = u_i,  \Lambda^{n+1}= \Lambda_i$.

\vspace{\baselineskip}

We now make a series of comments about the proposed approach.

\paragraph{Parallel implementation.} Step 1 is local, and can be solved trivially in parallel.  However, it is (generally) nonlinear and therefore requires an iterative approach.  In this work, we solve it using a steepest descent method with backtracking line search.   A potential difficulty is that different spatial points may require different number of iterations to converge, and we address it below.
Step 2 leads to a universal Poisson's equation for which there are a number of effective parallel solvers.  In this work, we consider problems with periodic boundary conditions and therefore use the fast Fourier transform.  Step 3 is a trivial local update, and step 4 a simple check.  Thus, this iterative algorithm can be implemented effectively using accelerators like Graphical Processing Units (GPUs) as we demonstrate in the subsequent sections.

\paragraph{Convergence.}

Boyd {\it et al.} \cite{boyd2011distributed} prove the convergence of the spatially discretized version of the algorithm for $\rho$ large enough under the hypotheses that $W$ is convex and the unaugmented functional with $\rho=0$ has a saddle point.   They also review improvements of this result in the literature.  However, it is not natural to explect $W$ to be convex in $F$ in finite deformation.  Here, there are results in the case of elasticity where there is no internal variable.  Glowinski and Le Tallec \cite{gl_siamjam_82} show that the weak form of equilibrium equation of incompressible elasticity is equivalent to the weak form of the first variation of the augmented Lagrangian functional.  Further, Glowinski and Le Tallec \cite{gl_siamjam_84} show in the case of Mooney-Rivlin materials that the finite element discretization of this iterative approach converges for sufficiently large $\rho$.  Furthermore, they show that the finite element solutions converge to the solution of the continuous problem.  Their arguments can be generalized to a larger class of incompressible, isotropic, polyconvex, hyperelastic materials \cite{glowinski1989augmented}.  However, the general case with internal variables remains open.

\paragraph{Connection to physics.} This iterative method also has a close connection to the physics.  Comparing ((\ref{eq:eq}), (\ref{eq:kin})) to ((\ref{eq:ss}), (\ref{eq:kr})), we see that step 1 is the constitutive update with the Lagrange multiplier converging to the stress.  Step 2 is the compatibility equation (\ref{eq:compat}) with the primal convergence in $r_p$.  Finally, dual convergence in  $r_d$ is equivalent to the equilibrium condition (\ref{eq:equil}) as shown in the appendix.

\paragraph{Penalty parameter.} 
We have noted above that the method is known to converge under suitable hypothesis on $W,D$ for all $\rho$ sufficiently large \cite{boyd2011distributed,gl_siamjam_84}.  However, the rate of convergence depends critically on the choice of $\rho$. We adapt an idea proposed by He, Yang and Wang~\cite{he2000alternating} to adaptively change $\rho$ with iteration guided by $r_p$ and $r_d$. A large $\rho$ better enforces  compatibility (cf. \ref{eq:helm}) and leads to a faster drop of the primal error $r_p$.  However, it leads to a poor enforcement of the constitutive equation (cf. \ref{eq:eq}) and slower drop of the dual error $r_d$. Conversely, a small $\rho$ leads to a faster drop of the dual error $r_d$, but a slower drop of the primal error $r_p$. After checking for convergence, we adjust $\rho$ as
\begin{equation}
    \rho_{i+1}= 
    \begin{cases}
     \kappa \rho_i, & \text{if} \quad r_p/r_d > \tau \\
    \max \{ \rho_i / \kappa, \rho_\text{min} \}, & \text{if} \quad r_d/r_p > \tau\\
    \rho_i,& \text{if} \quad else
    \end{cases}
    \label{eqnrho}
\end{equation}
for given $\kappa$, $\tau$ $\rho_\text{min}$.  We take $\kappa$ and $\tau$ to be $1.3$ and $10$ respectively to avoid too frequent updates.  $\rho_\text{min}$ enforces the requirement that $\rho$ remains large enough for the method to converge and the choice of $\rho_\text{min}$ depends on $W$.  An important observation is that this is enabled by the fact that the Laplace operator in (\ref{eq:helm}) is independent of $\rho$.

\paragraph{Approximate solution of the local problem.}
While the local problem  (\ref{eq:eq}), (\ref{eq:kin}) is trivially parallel, a potential problem is that different points may require a different number of iterations to converge to a given point-wise residual error.  Indeed, in practice (as we shall see in Section \ref{sec:num}), a few isolated points require a very large number of iterations to converge compared to the rest.  Unfortunately, the calculation can not advance to the next step till the last point has converged, and consequently, these slowly-converging points can add significantly to the computational cost.  However, Eckstein and Bertsekas~\cite{eckstein1992douglas} have proved that for convex $W$, the algorithm converges when the  the local error is summable.   In other words, it is not necessary to solve the local equations  (\ref{eq:eq}), (\ref{eq:kin}) at every point (i.e., in $L^\infty$ norm), but it suffices to solve them in some $L^p$ norm for appropriate $2 \le p < \infty$.    Working in $p=2$, set the local  residual to be 
\begin{equation}
    r_l = \frac{1}{\mu} || W_F(F^{n+1}, \lambda^{n+1}, x) - \Lambda^n + \rho( \nabla u^n - F^{n+1}) ||^2_{L^2}.
\end{equation}
We see from (\ref{eq:inexact}) in the Appendix that equilibrium is still satisfied if the both the dual and local residual go to zero.  Thus, satisfaction of the equilibrium equation does not require pointwise convergence.  Further savings can be achieved by keeping the local residual large in the initial (global) iterations, and gradually reducing it as (global) convergence is achieved \cite{boyd2011distributed}. We explore two strategies in Section \ref{sec:num}. In one, we maintain a balance between $r_l$ and $r_d$ while in the other we require a fixed fraction of local points to converge.  

While these approaches reduce the number of iterations of the local problem, computing either the local residual $r_l$ or the fraction of converged points requires a communication between the various points which can be expensive in an accelerator.  Therefore it is necessary to balance the cost of iteration and the cost of checking  convergence.  We study this balance in Section \ref{sec:num}.
%

\paragraph{Gradient internal variables.} 
In certain problems like phase transitions and microstructure evolution (like the one we shall study in Section \ref{sec:lce}), the state of the material is described not only by an internal variable, but also the gradient of the internal variable, i.e,. $W = W (F, \lambda, \nabla \lambda, x)$.  These can be incorporated into the method in two ways.  

The first approach is to treat the gradient of the internal variable in much the same way that we treat the deformation gradient.  But this requires some care to maintain the linear structure of step 2.  We introduce two internal variables, $\ell$ and $H$ and enforce the constraints $\ell = \lambda$, $H= \nabla \lambda$ using the augmented Lagrangian, i.e., consider the Lagrangian density
$W(F,\ell,H,x) + \Delta t D + \Lambda \cdot (\nabla u - F) + \rho/2 |\nabla u - F|^2 + \mu \cdot(\nabla \lambda - H) + \xi/2 |\nabla \lambda - H|^2 + \kappa \cdot (\ell - \lambda) + 
\zeta/2 |\ell - \lambda|^2$
with additional Lagrange multipliers $\mu, \kappa $ and penalty parameters $\xi, \zeta$.  We minimize the Lagrangian with respect to $F, \ell$ and $H$ in step 1 and with respect to $u$ and $\lambda$ in step 2.   We then update all the Lagrange multipliers in step 3 and check convergences in step 4.   Note that the equation in step 2 describing $\lambda$ is not the Poisson's equation but includes a linear term in $\lambda$; still, it can be treated as before, similar to Section \ref{sec:bloch}.

The second approach is to use the value of the gradient from the previous iteration.  In other words, we rewrite (\ref{eq:kin}) as 
\begin{equation} \label{eq:kin2}
W_\lambda (F_{i+1},\lambda_{i+1}, \nabla \lambda_i, x) + D_v \left( {\lambda_{i+1} - \lambda^n \over \Delta t} ,x \right) = 0 .
\end{equation}
While this is approximate, it is effective.  In most models, the gradient of the internal variable is introduced as a way of introducing a length scale, and this approximation does so effectively as we shall see in Section \ref{sec:lce}.

\subsection{Connection to other methods.}

\paragraph{Hu-Washizu variational principle.} 

The Hu-Washizu and other mixed methods have been used widely in mechanics (e.g. \cite{w_book_68}).  In the Hu-Washizu variational principle of finite elasticity, one seeks the stationary point of 
\begin{equation}
\int_\Omega \left( W(F) - S \cdot F - u \cdot (\nabla \cdot S) \right) dx
\end{equation}
(up to boundary terms) over $F, S, u$.  Integrating it by parts, we see that the functional is equivalent to 
\begin{equation}
\int_\Omega \left( W(F) - S \cdot (F - \nabla u) \right) dx
\end{equation}
(up to boundary terms).  This is the same as (\ref{eq:al}) with $\rho=0$ (and $S=\Lambda$).  Thus, our approach with $\rho=0$ reduces to the Hu-Washizu variational principle.  It is well-known that such variational principles are delicate with regard to convergence which we avoid with $\rho>0$.

\paragraph{Fast Fourier Transform-based methods.} 

Since the seminal work of Moulinec and Suquet \cite{moulinec1994fast} in the context of linear composite materials, a number of FFT-based methods have been used to study various problems of micromechanics as described in Section \ref{sec:intro}.   While the details differ in the various authors, the basic method involves solving the equilibrium equation using FFT.  Linearize (\ref{eq:elas}) around $u_0$ to obtain
\begin{equation}
- \nabla \cdot {\mathbb C}(x) \nabla w = f
\end{equation}
where ${\mathbb C}(x) =   W_{FF} (\nabla u_0(x), \lambda (x), x)  $ and $f = \nabla \cdot W_{F} (\nabla u_0(x), \lambda (x), x)$ and $w=u-u_0$.  We introduce a homogenous comparison medium ${\mathbb C}_0$ and rewrite the above as
\begin{equation}
- \nabla \cdot {\mathbb C}_0 \nabla w = f +\nabla \cdot ({\mathbb C}- {\mathbb C}_0) \nabla w 
\end{equation}
which is a Lippmann-Schwinger type equation.  We solve this iteratively as
\begin{equation}
- \nabla \cdot {\mathbb C}_0 \nabla w^{n+1} = f +\nabla \cdot ({\mathbb C}- {\mathbb C}_0) \nabla w^n.
\end{equation}
using FFT since it has homogeneous coefficients. The convergence as well as the rate of convergence depends sensitively on the choice of the comparison medium ${\mathbb C}_0$, especially when the `contrast' (i.e., the range of ${\mathbb C}$) is large.  This has led to formulations beyond the Lippmann-Schwinger equation \cite{eyre1999fast, michel2000computational, monchiet2012polarization} (we refer the reader to \cite{moulinec2014comparison,msm_ijnme_18} for a further discussion of these methods) through the introduction of operators that enable fast convergence.

\paragraph{Augmented-Lagrangian method.}

Notable amongst FFT-based methods is the augmented Lagrangian method of Michel, Moulinec and Suquet \cite{michel2000computational}.  In our language, they study the functional
\begin{equation} \label{eq:almms}
\int_\Omega \left( W( F, \lambda, x) + \Delta t  D\left( {\lambda - \lambda^n \over \Delta t},x \right) + \Lambda \cdot (\nabla u - F) +
{\rho \over 2} (\nabla u - F) \cdot {\mathbb C}_0 (\nabla u - F)   \right) dx
\end{equation}
where $\rho{\mathbb C}_0$ is the modulus of a comparison medium.  They study convergence with $\rho$ at low spatial resolution and use this value at high spatial resolution.  Our approach is a special case of their method with ${\mathbb C}_0 = $ Id.  The advantage of doing so is that it leads to a Helmholtz projection in Step 2 as opposed to one that depends on the comparison medium.  Further, we tune $\rho$ during iteration (cf. (\ref{eqnrho})).

This algorithm was proposed for finite elasticity (no internal variables) in the context of general boundary value problems and finite elements by Glowinski and Le Tallec   \cite{gl_siamjam_82,gl_siamjam_84}.

\paragraph{Split Bregman method.}  
The split Bregman method introduced by Goldstein and Osher \cite{go_siamjis_08} (based on an earlier method proposed by Bregman \cite{breg} to regularize convex problems) seeks to minimize a functional $J(F)$ amongst all minimizers of another functional $H(F)$.   In our case, we take these functionals to be
\begin{equation}
J(F) = \int_\Omega \left( W( F, \lambda, x) + \Delta t  D\left( {\lambda - \lambda^n \over \Delta t},x \right) \right)dx, \quad
H(F) = \int_\Omega  |\nabla u - F|^2 dx.
\end{equation}
It has been proven to be equivalent to the augmented Lagrangian approach \cite{yin} and used in a variety of problems including image sensing \cite{go_siamjis_08,yin}, free boundary problems \cite{gu_jcp_20} and microstructure formation \cite{jv}.

\subsection{Implementation in the periodic setting}

It is common in computational micromechanics to consider a material that is periodic and a representative volume element that is a unit cell.  In other words, $\Omega = (-L,L)^3$ and $W(F,\lambda,x), D(\lambda,x)$ are periodic in $x$.  Further, the average deformation gradient $\langle F \rangle$ is prescribed in strain-controlled simulations, the average stress $\langle S \rangle$ is prescribed in stress-controlled simulations and some  combination with some components of $\langle F \rangle$ and complementary components of $\langle S \rangle$ are prescribed in mixed simulations (e.g., plane stress where the average planar stretch is prescribed).   Above, $\langle \cdot \rangle$ denotes spatial average.

We assume that the resulting deformation gradient $F$, internal variable $\lambda$ and stress $S$ are also periodic in $x$.  This implies that the deformation $u$ is periodic up to a linear function; i.e., $u - \langle F \rangle x$ is periodic.  It is then natural to solve Step 2 (\ref{eq:helm}) using fast Fourier transforms (FFT).  In Fourier space, (\ref{eq:helm}) becomes local (i.e., can be solved at each $\xi$) as
\begin{equation} \label{eq:helmf}
\hat u (\xi) = - { (\hat F  - \rho^{-1} \hat \Lambda) i \xi \over |\xi|^2 }.
\end{equation}
where $\hat f (\xi) $ denotes the Fourier transform of $f(x)$ and $i$ is the imaginary unit.  Thus, we solve Steps 1, 3 and 4 in real space and Step 3 in Fourier space using FFT and iFFT (inverse fast Fourier transform) to go back and forth between them.  Specifically, we consider a regular $N \times N \times N$ grid (for $N$ even) on $\Omega$ in real space, and the corresponding $N \times N \times N$ on the domain $(-{\pi \over L}, {\pi \over L})^3$ in Fourier space.  We define  $\hat f (\xi) $ as the discrete Fourier transform on this space and use FFT to evaluate it.

A couple of comments are in order.  First, it is convenient to work with 
\begin{equation}
\tilde u = u - \langle \nabla u \rangle x, \quad \tilde F = F - \langle F \rangle, \quad \tilde \Lambda = \Lambda - \langle \Lambda \rangle
\end{equation}
which are all periodic (recall that $u$ is not necessarily periodic).  This is also convenient since the boundary conditions are prescribed in terms of $\langle F \rangle$ or  $\langle S \rangle$  (recall $\Lambda$ converges to $S$).

Second, the material may be heterogeneous and $W, D$ may be discontinuous functions of $x$ in many problems of interest.  In such situations, $F, \lambda$ may discontinuous and thus the use of Fourier transforms to solve for $u$ may lead to spurious oscillations.  An approach around this that has proved very effective  \cite{berbenni2014numerical,lebensohn2012elasto,vidyasagar2017predicting} in various problems using the closely related Lippmann-Schwinger approach  is to replace the discrete Fourier transform of the derivatives with the discrete Fourier transform of the central differences:
\begin{align}
&\widehat{u_{i,j}}(\xi) = i \hat{u}_i (\xi)  \xi_j \quad \text{with} \quad
{\widehat{u_i(x+he_j)} - \widehat{u_i(x-he_j)} \over 2h}(\xi) = i \hat{u}_i (\xi) { \sin (h \xi_j) \over h }, \\
&\widehat{u_{i,jj}}(\xi) = - \hat{u}_i (\xi)  |\xi_j|^2 \quad \text{with} \quad
{\widehat{u_i(x+he_j)} + \widehat{u_i(x-he_j)} - 2\widehat{u_i(x)} \over h^2}(\xi) = - \hat{u}_i (\xi) { 4\sin^2 ({h \xi_j \over 2}) \over h^2 }.
\end{align}
Since $\sin \alpha / \alpha < 1$, this is equivalent to a high frequency filter and suppresses the spurious oscillations.  We refer the reader to \cite{zetal_inprep_20} for further discussion.

\subsection{Quadratic functionals and Bloch waves} \label{sec:bloch}

We conclude this section with the discussion of a closely-related linear problem that arises in the study of stability of periodic solutions to nonlinear problems.  Let ${\mathbb L}_{ijkl}(x)$ be a periodic fourth order tensor field of period 1 with ${\mathbb L}_{ijkl} = {\mathbb L}_{klij}$.  We are interested in evaluating
\begin{align}
    \beta_k = \min_{v \in {\mathcal A}_k}  \int_{\Omega_0} {1 \over 2} \overline{v}_{i,j} {\mathbb L}_{ijkl} v_{k,l} d x = 
    \min_{v \in {\mathcal A}_k}  \int_{\Omega_0} \nabla v \cdot {\mathbb L} \nabla v \ d x
    \label{eqnbeta}
\end{align}
over an admissible class of functions that are unit Bloch waves:
\begin{align}
{\mathcal A}_k= \left\{v \in H^1(\Omega_0): ||v||_{L^2 (\Omega_0)} = 1, \ v(x) = p(x) \exp(i \omega_k \cdot x), \ 
\omega_k = \left\{ {2\pi \over k_1},  {2\pi \over k_2} \right\}, \ p \ 1-\text{periodic} \right\} 
\end{align}
for $k= \{k_1, k_2\}$ with $k_i$ integers.  Above, $\Omega_0 = (0,1)^2$ is the unit square, and $\bar{v}$ denotes the complex conjugate of $v$.
As before, we use the augmented Lagrangian formulation to write
\begin{align}
    \beta_k = \min_{v \in {\mathcal A}_k, F\in L^2(\Omega_0)} \ \max_{\Lambda \in L^2(\Omega_0)} 
    \int_{\Omega_0} \left( {1 \over 2} F \cdot  \mathbb{L} F + \Lambda \cdot (\nabla v - F) + {\rho \over 2} |\nabla v - F|^2 \right)dx.
\end{align}
Recalling that $v$ is a Bloch wave, and setting $F(x) = G(x) exp(i \omega_k \cdot x)$, $\Lambda(x) = g(x) exp(i \omega_k \cdot x)$,
where $G, g \in L^2(\Omega_0)$ extended periodically, it follows that 
\begin{align}
    \beta_k = \min_{p \in {\mathcal P}, G \in L^(\Omega_0) } \max_{g \in L^(\Omega_0) }
        \int_{\Omega_0} \left( {1 \over 2} G \cdot  \mathbb{L} G + g \cdot ((i \omega_k+\nabla) p - G) + {\rho \over 2} | (i \omega_k+\nabla) p - G|^2 \right)dx
\end{align}
where ${\mathcal P} = \{ p \in H^1(\Omega_0): ||p||_{L^2(\Omega_0)} = 1, p \ 1-\text{periodic} \}$.  This saddle point problem can be solved as before using ADMM.  Given $G_i, p_i, g_i$,
\begin{itemize}
\item {\it Step 1': Local problem.} Update $G$: $G_i = (\mathbb{L} + \rho \mathbb I)^{-1} (g_i + \rho (i \omega_k + \nabla) p^n)$;
\item {\it Step 2': Global update.} Update $p$: $(i\omega_k + \nabla)^2 p_{i+1} = (i \omega_k + \nabla) \cdot (G_{i+1} - \rho^{-1} g_i)$;
\item {\it Step 3': Update Lagrange multiplier.}  Update $g$:  $g_{i+1} = g_i + \rho ( (i \omega_k + \nabla)p_{i+1} - G_{i+1})$;
\item{\it Step 4': Check for convergence}.
\end{itemize}
Note that the global problem can be solved trivially in Fourier space.  

Thus, a quadratic functional can be minimized over Bloch waves in the original unit cell with a slight modification of our algorithm.

\section{GPU implementation} \label{sec:imp}

\begin{figure}
    \centering
    \includegraphics[width=0.6\textwidth]{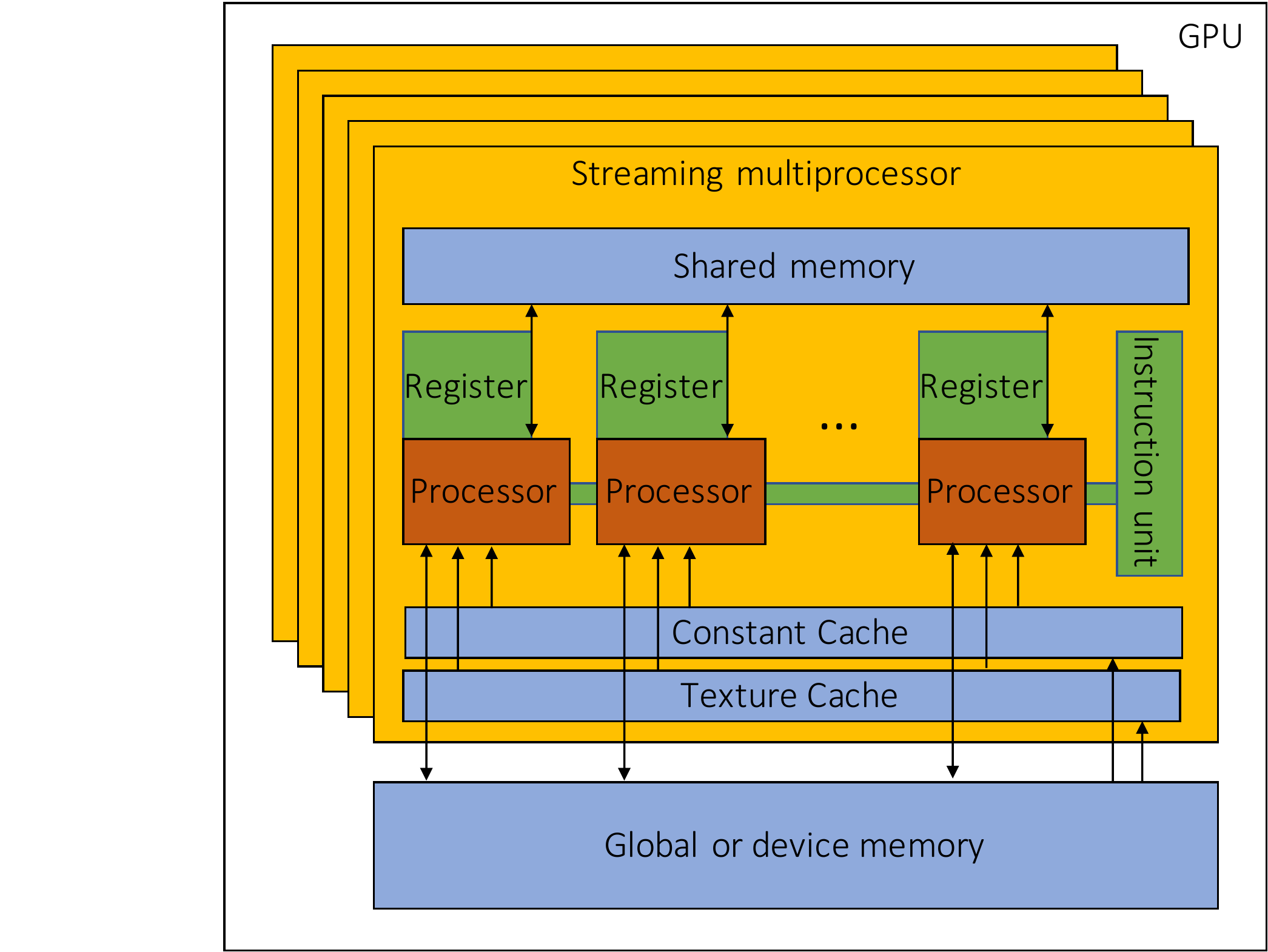}
    \caption{A schematic representation of the architecture of a general purpose graphical processing unit (adapted from \cite{preis2009gpu})}
    \label{Neo_config}
\end{figure}

We begin with a brief introduction to general purpose GPUs and their use in computing, referring the reader to \cite{kirkhwu} for details.
A compute node typically consists of a single CPU and one or more general purpose GPUs.   While the clock speed of a CPU is faster than that of the GPU, the presence of thousands of cores and the architecture enables faster overall performance if properly organized.  
A schematic representation of a general purpose GPU is shown in Figure \ref{Neo_config}.   It consists of a global or device memory and a number of streaming multiprocessors (SMs).  Each SM in turn contains a number of cores or processors that have access to a shared memory, various registers and an instruction unit.  All SMs also have access to the constant cache and the texture cache.   The calculation is organized in threads with each processor typically executing a single thread.  The threads are organized into warps.  All threads in the warp work following the ``single instruction multiple data (SIMD)'' organization, i.e. each processor executes the same instruction concurrently with possibly different data.  IF there are many conditional instructions and different threads fall into different conditions, then it leads to a situation described as `warp divergence' where each of the conditional instructions is executed in a serial manner.  It is important to avoid warp divergence.  

The exchange of data between a CPU and a GPU is slow, and therefore has to be minimized.  Even though significantly faster, the exchange of data within a GPU between the global memory and the shared memory of a SM is also slow.  However, this can be made faster using a parallel transfer strategy called `coalesced memory access' \cite{wu2013complexity} when  a one to one mapping can be created between a thread and a segment of the global memory.

\begin{algorithm}[t]
\SetAlgoLined
    Given an initial value of $u^0, F^0, \lambda^0$ and macroscopic strain path $\overline{F}(t)$\;
    {\it Step 0: Initialize}:\\
    	\ \ \ \ Place $\rho, \epsilon_p, \epsilon_d, \gamma_p, \gamma_d, f$ in constant cache;\\
    	\ \ \ \ Place $u^0, F^0, \lambda^0, \Lambda^0$ in global memory,\\
    \While{$t < t_\text{max}$}{
    Initialize $F_0 = F^n, \lambda_0 = \lambda^n, u_0 = u^n, \Lambda_0 = \Lambda^n$\\
    \While{$r_p > \epsilon_p$ \text{or} $r_d > \epsilon_d $}{
         {\it Step 1: Local problem.} \\
                 \While{$r_l > f r_d$}{
	         	Using kernel function:\\
        			\ \ \ \ $\cdot$ Move $F_i, u_i, \lambda_i, \Lambda_i$ to shared memory;\\
        			\ \ \ \ $\cdot$ Fixed number of iterations to solve (\ref{eq:eq}), (\ref{eq:kin}) for $F_{i+1}, \lambda_{i+1}$;\\
        			\ \ \ \ $\cdot$ Move $F_{i+1}, \lambda_{i+1}$ to global memory\\
			Compute $r_l$  \tcc*{ \texttt{cuBlas}}}
       {\it Step 2: Helmholtz projection.} \\
         	\ \ \ \ $\cdot$ FFT  $\widehat{F}_{i+1}$, $\widehat{\lambda}_{i+1}$ \tcc*{ \texttt{cuFFT}}
		\ \ \ \ $\cdot$ Using kernel function:\\
		\ \ \ \ \ \ \ \ $\cdot$ Move $\widehat{F}_{i+1}$, $\widehat{\lambda}_{i+1}$ to shared memory;\\
		\ \ \ \ \ \ \ \ $\cdot$ Find $\widehat{u}_{i+1}$ from (\ref{eq:helmf}); \\
		\ \ \ \ \ \ \ \ $\cdot$ Move $\widehat{u}_{i+1}$ to global memory; \\
		\ \ \ \ $\cdot$ iFFT $\widehat{u}_{i+1}$ \tcc*{ \texttt{cuFFT}}
         {\it Step 3: Update Lagrange multiplier.} Find $\Lambda_{i+1}$ from (\ref{eq:update})    \tcc*{ \texttt{cuBlas}}
         {\it Step 4: Check for convergence}  Compute $r_p, r_d$ from (\ref{eq:check}) 		   \tcc*{ \texttt{cuBlas}}
    }
    Update $t$, $F^{n+1},u^{n+1}, \lambda^{n+1}, \Lambda^{n+1}$
    }
\caption{Implementation on GPUs \label{alg}}
\end{algorithm}

The implementation of the algorithm described earlier is presented in Algorithm \ref{alg}.  We work on a compute node consisting of a 14-core Intel Broadwell CPU and four Nvidia Tesla P100 GPUs.  Each GPU contains 16GB of (global or device) memory and 56 SMs with 64 cores each for a total of 3584 cores, and has a double precision performance of 4.7 teraFLOPS.   The P100 GPUs enable the compute unified device architecture (CUDA) platform with the standard programming language C++.  CUDA uses warps of 32 threads  \cite{kirkhwu}.  The implementation is general and can be expanded to other platforms.

The algorithm takes advantage of the GPU architecture in various ways:
\begin{itemize}
\item All calculations are performed on the GPU.  The CPU is used only for kernal function calls (i.e., to provide instructions to the GPU), initialization and output of results.
\item The exchange of data between CPU and GPU is limited to the first initialization and to write results.
\item Following first initialization, all data is kept on the GPU global memory during the entire calculation.   Since the proposed algorithm uses the result of the previous time step to initialize the current time step, it is not necessary to perform any GPU/CPU transfer between time steps when there is no need to write the result.
\item Global constants like the penalty and tolerance parameters are kept in constant cache.
\item The local step 1 of the proposed algorithm is well-suited for SIMD since the same equations are solved independently at each point.  Further, the structure  enables optimization of the exchange between global and shared memory within the GPU in two ways.  The data can be kept in shared memory within the SM during the local iterations.   Therefore, the exchange between global and shared memory is limited to the initialization and to the final output of the local iteration.  Even these transfers can exploit the coalesced memory access since each thread (point) only requires data stored at a particular location in global memory. 
\item The Helmholtz projection is local in Fourier space (cf. (\ref{eq:helmf})).  Therefore is it well-suited for SIMD, and can take advantage of  coalesced memory access.
\item The local nature of the local step 1 in real space and the Helmholtz projection in Fourier space avoid warp divergence.
\item The computation of the  L2 norms in the approximate solution of step 1, the Lagrange muliplier update (step 3), the convergence check (step 4) are executed using basic linear algebra operations.
\item Library functions that are optimized for GPUs are available for fast Fourier transform and basic linear algebra operations.
\end{itemize}

\section{Bifurcation in finite elasticity} \label{sec:bif}

We start with a problem of finite deformation that involves a bifurcation, and one that has been previously studied using both computation in two dimensions \cite{triantafyllidis2006failure,bertoldi2008mechanics} and experiment \cite{mullin2007pattern,bertoldi2008mechanics}.  It serves as verification of the proposed method and algorithm against previous results of Triantafyllidis, Nesterovi{\'c} and Schraad \cite{triantafyllidis2006failure}, Section \ref{sec:bi2d}.  We also use this example to discuss convergence and scaling in Section \ref{sec:num}.

\subsection{Bifurcation} \label{sec:bi2d}

\begin{figure}[t] 
\centering
\includegraphics[width=0.8\textwidth]{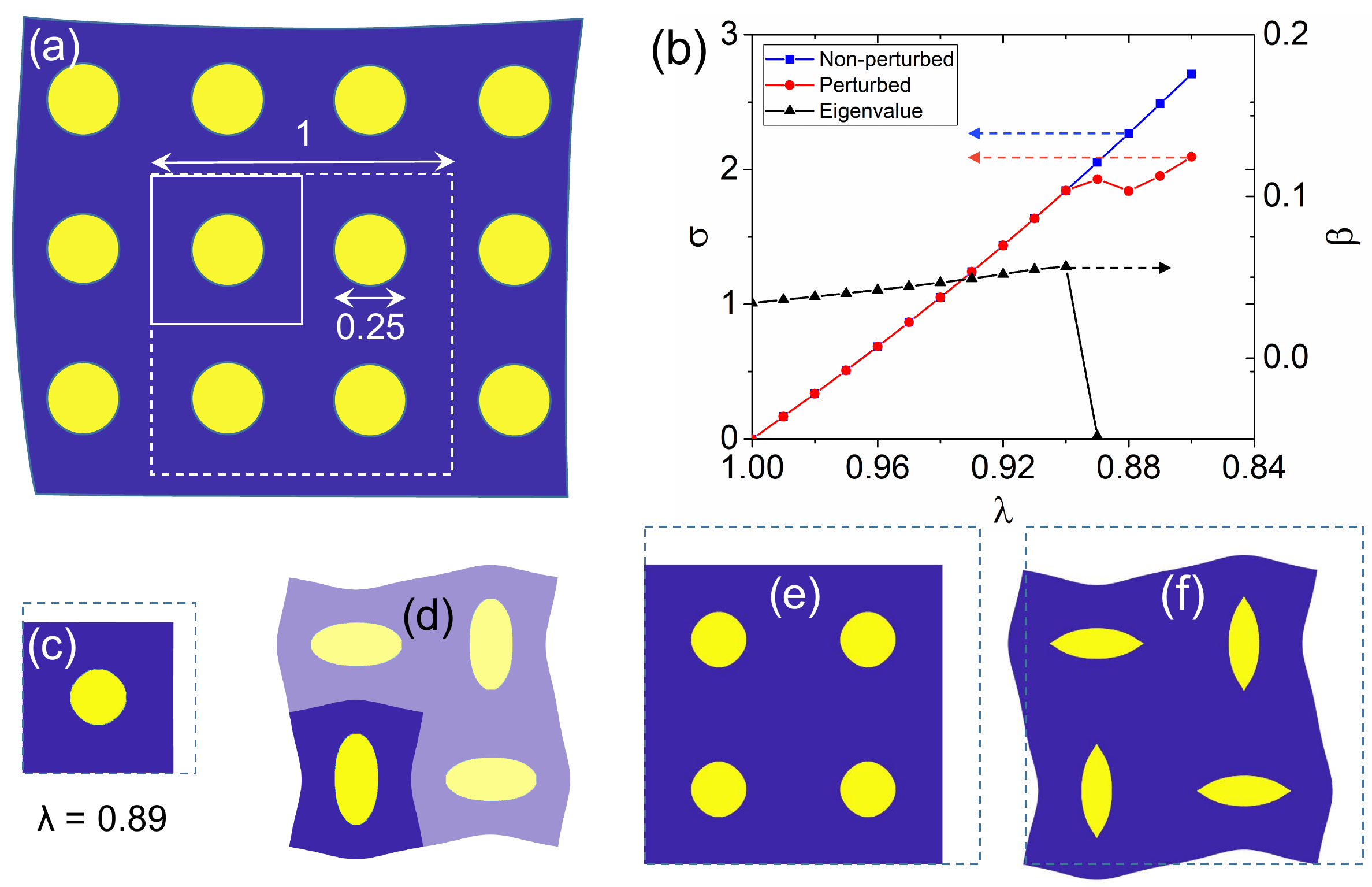}
\caption{Bifurcation of a periodic composite.  (a) A periodic composite with the unit cell (bold lines) and super cell (dashed line).  (b) The stress-stretch curve without bifurcation (blue, square symbols) and with bifurcation (red, round symbols) along with the modulus of stability $\beta_{(2,2)}$ (black, triangle symbols).  (c) Deformed shape of the unit cell (dashed line showing the undeformed size) at $\lambda =0.89$). (d) The mode shape of the unstable mode in the period-doubling instability at $\lambda =0.89$. (e) The deformed shape of the supercell without bifurcation at $\lambda =0.89$.  (f) The deformed shape of the supercell post bifurcation at $\lambda =0.89$.  \label{fig:bif} }
\end{figure}

Consider a periodic arrangement of compliant circular inclusions in a stiff matrix in two dimensions as shown in Figure \ref{fig:bif}(a).  Both materials are modeled as compressible Mooney-Rivlin materials with stored energy density
\begin{equation}
    W(F,x) = \frac{\mu(x)}{2} (I_1 - \ln{I_2} -2) + \frac{\kappa(x)}{2} (\sqrt{I_2} - 1)^2
\end{equation}
where $\mu$ is the shear modulus and $\kappa$ is the bulk modulus, and $I_1$ and $I_2$ are the first and second invariants of right Cauchy-Green tensor $C$.  The moduli take the values $\mu_i, \kappa_i$  and $\mu_m, \kappa_m$ in the inclusion and the matrix respectively with $\mu_i < \mu_m, \kappa_i < \kappa_m$.   In our numerical examples, $\kappa_i/\mu_i = \kappa_i/\mu_m = 9.8$ while $\mu_m/\mu_i =20$.

We start with a $1 \times 1$ unit cell simulation. The periodic medium is subjected to equi-biaxial compression, $\langle F \rangle = \lambda I$  where $\lambda$ decreases monotonically from an initial value of 1.  At each given value of $\lambda$, the equilibrium solution is computed using Algorithm 1 on a $1024 \times 1024$ grid starting with the solution of the previous $\lambda$ as an initial guess.  We obtain the stress-stretch curve shown by square symbols (blue) in Figure \ref{fig:bif}(b) and a periodic solution with the deformed unit cell shown in Figure \ref{fig:bif}(c).

It is known that this example develops a long-wavelength instability.  Note that a 1-periodic function is also $k-$periodic for any $k=(k_1,k_2), k_i$ integers.  Thus, we may have equilibrium solutions that are periodic on a $k_1 \times k_2$ super-cell.  However, it is known from 
Geymonat, M\"uller and Triantafyllidis \cite{gmt_arma_93} that the 1-periodic solution is the stable solution near $\lambda =1$.  However, this solution may become unstable as $\lambda$ changes.  By the second variation condition, the 1-periodic solution $u$ remains stable as long as 
\begin{equation}
\int_{\Omega_k} \nabla v \cdot {\partial W^2 \over \partial F \partial F} (\nabla u(x),x) \nabla v dx \ge 0
\end{equation}
for all non-zero $k-$periodic functions $v$.  Using Bloch waves, this is equivalent to requiring $\beta_k \ge 0$ where $\beta_k$ is as defined in (\ref{eqnbeta}).  

Therefore, we compute the modulus of stability $\beta_k$ for various $k$, and this is also shown with triangular symbols (black) in Figure \ref{fig:bif}(b).  We see that $\beta_{(2,2)} \to 0$ as $\lambda \to 0.9$.  The corresponding mode $v_k$ is shown in Figure \ref{fig:bif}(d) as the darkened region.  This suggests that the periodic solution will bifurcate to a solution that is periodic on a $2 \times 2$ super-cell.   

We therefore repeat the finite deformation equilibrium computation on a $2 \times 2$ super-cell and a $2048 \times 2048$ mesh, once without a perturbation (i.e., with the solution to the previous $\lambda$ as an initial guess), and once with the linearly unstable mode added as a perturbation (i.e., with the sum of the 
solution to the previous $\lambda$ and a scaled eigenmode $v_k$ associated with $\beta_k$  as the initial guess).  The simulation without a perturbation leads to a periodic solution as before Figure \ref{fig:bif}(e) and with the same stress-stretch curve shown in black in Figure \ref{fig:bif}(b).   The perturbed solution also agrees with it until $\lambda \approx 0.9$, but then bifurcates into a solution with period $2 \times 2$ shown in Figure \ref{fig:bif}(f) with a stress-stretch curve shown with round symbols (red).

All results agree with those of Triantafyllidis, Nesterovi{\'c} and Schraad  \cite{triantafyllidis2006failure}, thereby verifying the method.

\subsection{Convergence and performance} \label{sec:num}

We now use this example to demonstrate convergence and scaling of the proposed algorithm.  In all the tests, we compress the composite until $\lambda = 0.95$.

\begin{figure}
\centering
\includegraphics[width=0.9\textwidth]{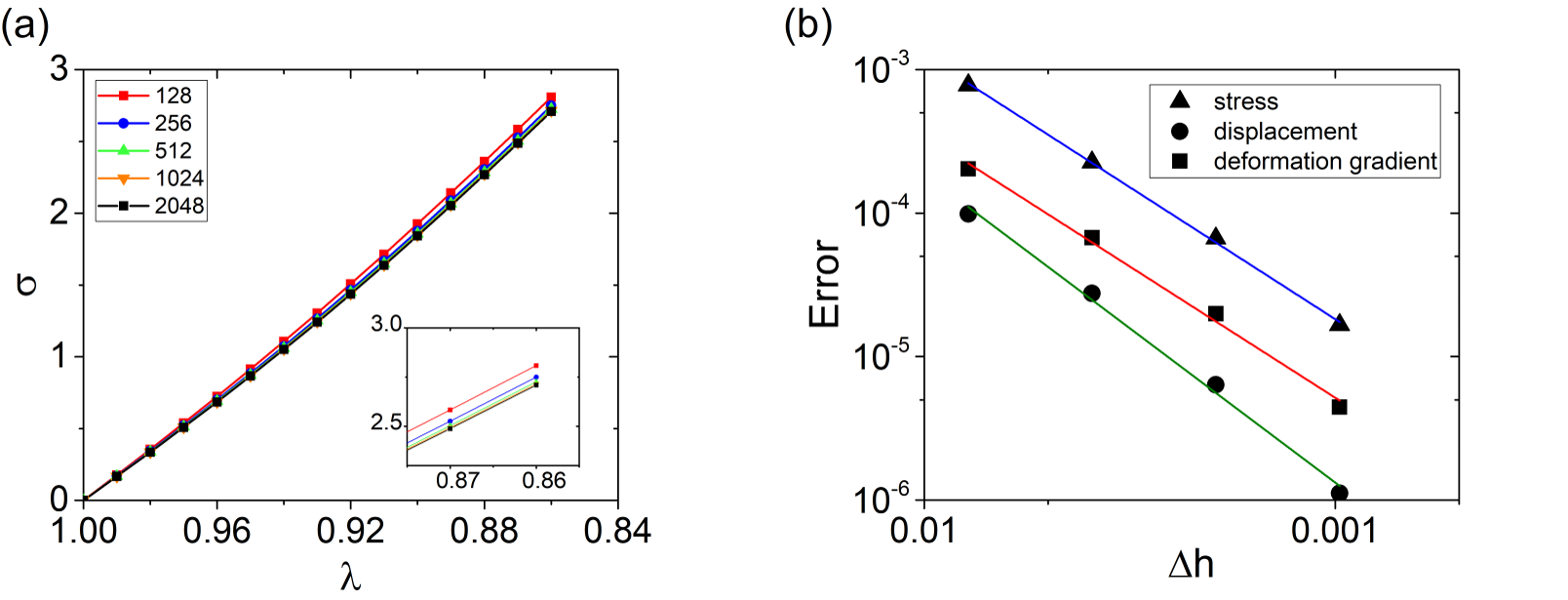}
\caption{Convergence with mesh size. (a) Stress-stretch curve for various computational grid resolution and  (b) Relative error in the deformation gradient and stress vs. grid size.}
\label{fig:conv}
\end{figure}

We begin by investigating the convergence with resolution. The simulation is performed with $128 \times 128$, $256 \times 256$, $512 \times 512$, $1024 \times 1024$, and $2048 \times 2048$ grids, and the stress-stretch curves are shown in Figure \ref{fig:conv}(a).  Further, taking the $2048 \times 2048$ grid as the reference, Figure \ref{fig:conv}(b) shows how the relative error ($L_2$ norm) of deformation gradient and stress depends on resolution.  We observe polynomial convergence with rates of $1.83$ and $1.84$ for the error in deformation gradient and displacement respectively. These are very close to the expected rate of $2$ for the discrete differential operator of FFT.  We believe that these are due to the change in residual spurious oscillation at the interface as well as the change in pixellated geometric representation with resolution.

\begin{figure}
\centering
\includegraphics[width=0.9\textwidth]{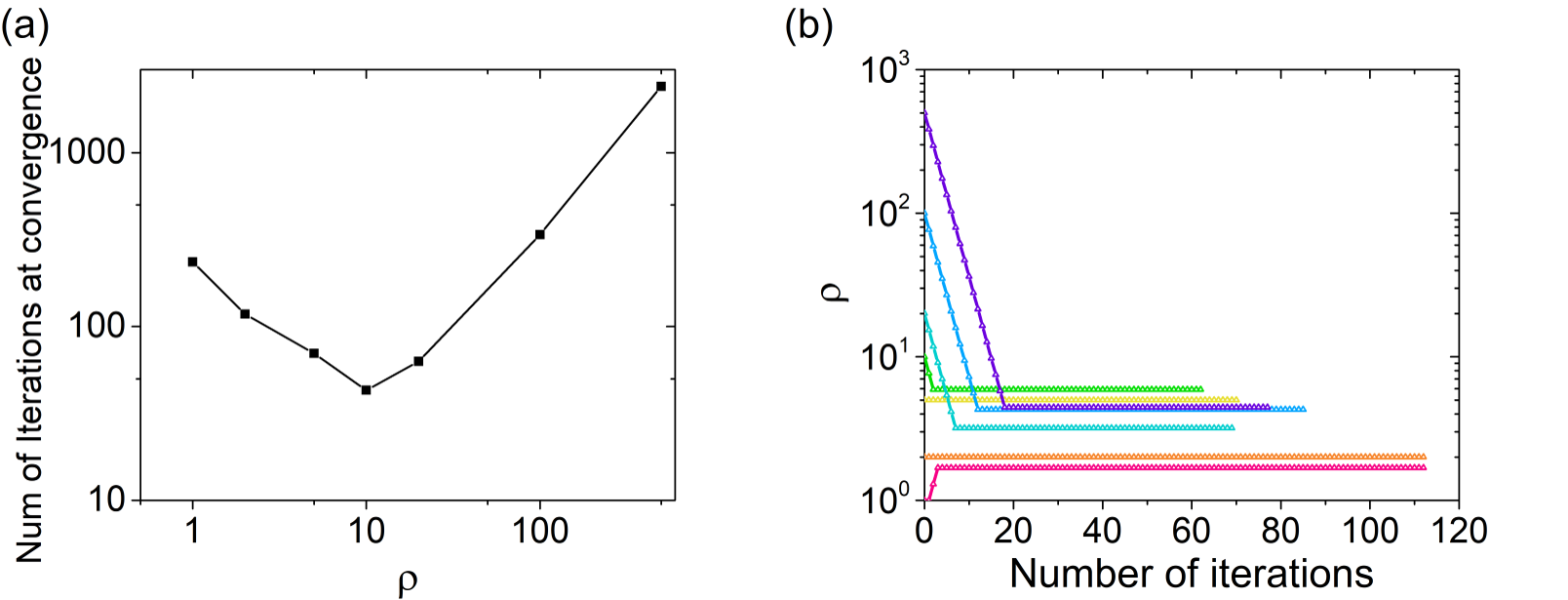}
\caption{Effect of penalization parameter $\rho$ on convergence.  (a) Total number of iterations with different fixed $\rho$ and (b)Variation of $\rho$ starting from different values.}
\label{fig:iter}
\end{figure}

The effect of the penalization parameter $\rho$ on the number of iterations required for a given error is shown in Figure \ref{fig:iter}.  
 Figure \ref{fig:iter}(a) shows the number of iterations to achieve a given convergence (in primal and dual error) when $\rho$ is held fixed at different values.  We observe that the number of iterations increase for both small $\rho$ and large $\rho$ with the optimal at about $\rho = 10$.   As noted earlier in Section \ref{sec:form}, the primal error is large for small $\rho$ and the dual error is large for large $\rho$.  This is the reason that we adjust $\rho$ following (\ref{eqnrho}). We show the evolution of $\rho$ for various initial values of $\rho$ in Figure \ref{fig:iter}(b) with $\kappa$ and $\tau$ in (\ref{eqnrho}) set to $5$ and $1.3$ respectively.  We note that in all cases, $\rho$ converges exponentially to values from $3$ to $10$.   Further, in contrast to the case with fixed $\rho$, the simulation converges well before $100$ iterations.  Thus, (\ref{eqnrho}) ensures a robust convergence of $\rho$ and significantly speeds up the algorithm. We have also observed  in our numerical experiments that $\rho < 1$  leads to divergence early in the iteration.

\begin{figure}[t!]
\centering
\includegraphics[width=0.9\textwidth]{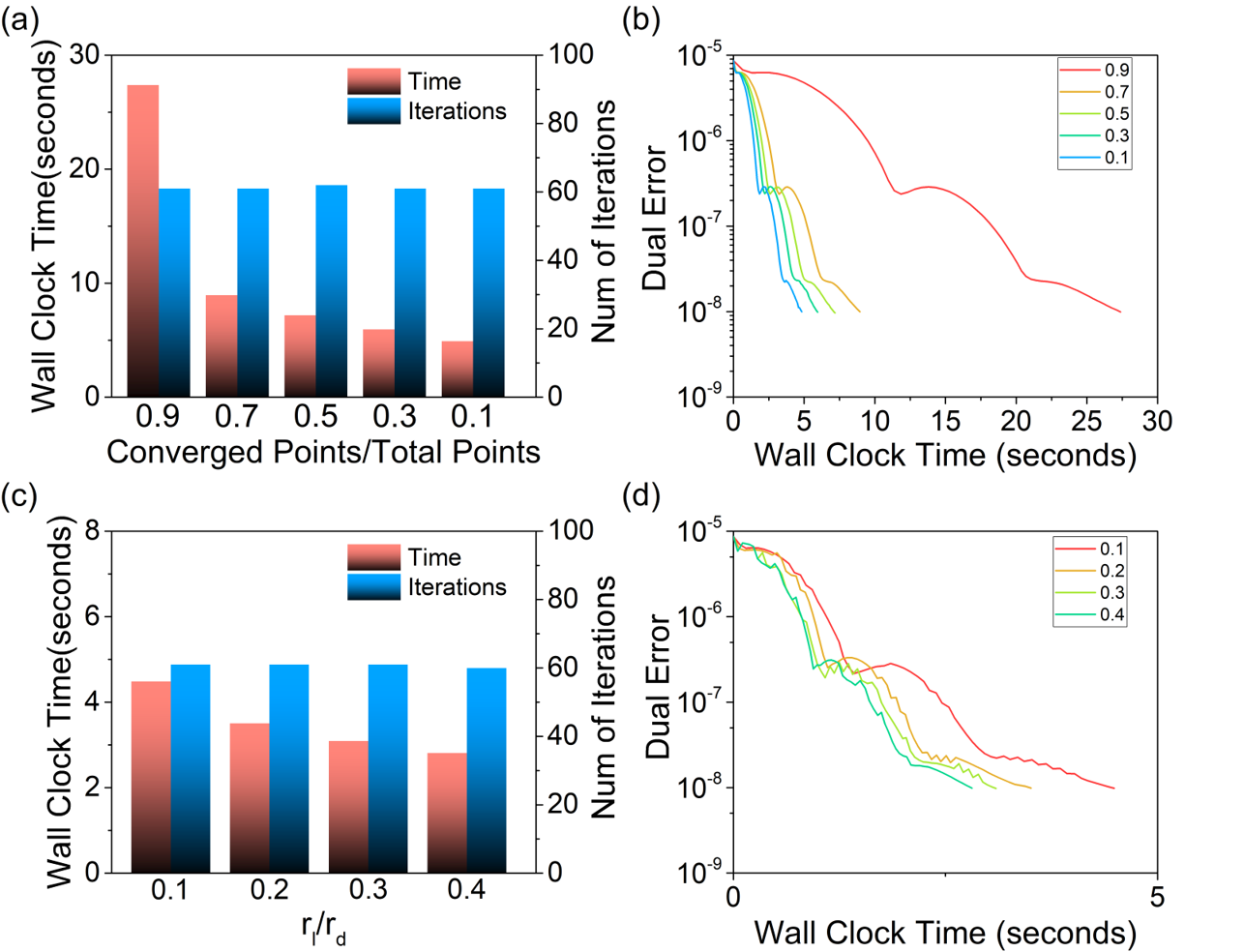}
\caption{Performance with approximate solution of the local problem.  (a,b) Local convergence on a fixed fraction of spatial points: (a) Wall clock time and number of global iterations for global convergence for various fractions.  (b)  The global dual error versus wall clock time for various fractions.  (c,d) Fixed ratio of local ($r_l$) to global dual ($r_d$) residual: (c) Wall clock time and number of global iterations for global convergence for various ratios and (d) The global dual error versus wall clock time for various ratios.}
\label{fig:inexact}
\end{figure}


We study the approximate solution of the local step 1 in Figure \ref{fig:inexact} using two strategies.  The first strategy is to require convergence of the local iterations of Step 1 only on a fixed fraction of spatial points, and these results are shown in Figure  \ref{fig:inexact}(a,b).  We check how many spatial points have converged to a given (pointwise) residual after a given number of local iterations and proceed to step 2 if a given percentage of spatial points have converged.  This check requires a communication between the shared and device memory which adds time, but it can be expedited using coalesced memory (the time required for the memory transfer and check is comparable to the time required for a single local iteration in our examples).  Figure \ref{fig:inexact}(a,b) shows the results when the local residual is held to $10^{-11}$, a check is performed every two local iterations and the percentage of converged points varied from 10\% to 90\%.   We see from Figure \ref{fig:inexact}(a) that the wall clock time for the global iteration to converge decreases monotonically with the percentage of converged points.  Further, the number of global iterations necessary for global convergence is largely independent of the number of converged points.  In other words, the approximate solution of the local step 1 has relatively little adverse effect on the global convergence.    Indeed, we see from 
Figure \ref{fig:inexact}(b) that the global dual residual decreases in the same manner as the calculations proceed, except  each global iteration is faster 
thereby reducing the overall clock time.

The second strategy is to require that the local residual $r_l$ be a fixed fraction of the global dual residual $r_d$, and the results are shown in Figure \ref{fig:inexact}(c,d).  We observe from Figure \ref{fig:inexact}(c) that the wall clock time for global convergence decreases while the total number of global iterations necessary for global convergence remains unchanged as we increase the ratio $r_l/r_d$.  In fact, at around a ratio of 0.3, we only require one or two iterations in the local step after a few global steps.  Again, we see from Figure \ref{fig:inexact}(d) that the global dual residual decreases in the same manner independent of the ratio $r_l/r_d$ as the calculations proceed, except  each global iteration is faster with increasing $r_l/r_d$ thereby reducing the overall clock time.    

These results in Figure \ref{fig:inexact} show that approximate solution of the local step 1 is an effective strategy to improving performance of our method.  In effect, we allow for larger tolerance in the local step when the global residual is large and exploit these in future iterations.  Additionally, recall from Section \ref{sec:form} that the error in satisfying the (physical) equilibrium equation is bounded by the local and residual global error.  Therefore, requiring the local residual to be a fraction of the global residual ensures physically meaningful solutions.  Therefore, we adopt this strategy.  Finally, we remark that this strategy is especially useful in highly nonlinear problems.  The analogous results for our example in liquid crystal elastomers are shown in Figure \ref{fig:lceinexact} of the Appendix.

\begin{figure}
\centering
\includegraphics[width=0.9\textwidth]{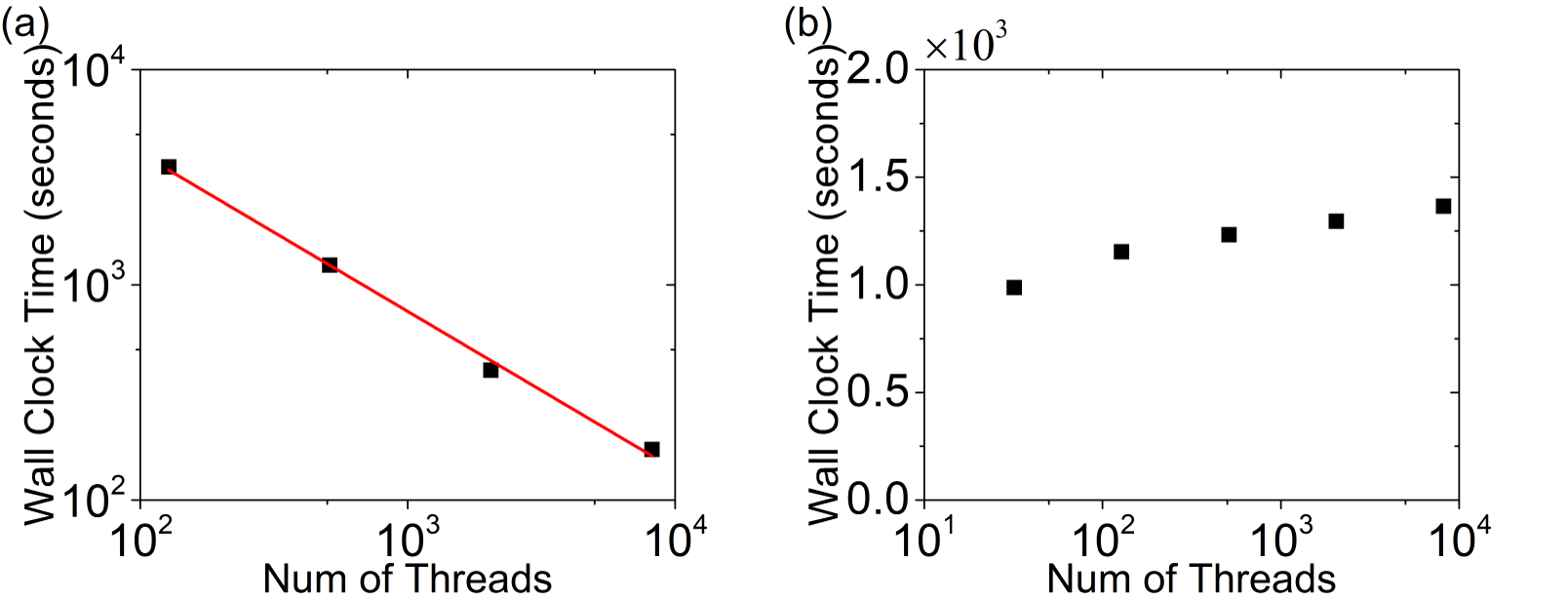}
\caption{Parallel performance of the algorithm.  (a) Strong scaling and (b) Weak scaling.}
\label{fig:scale}
\end{figure}

Finally, the parallel performance of the algorithm is examined in Figure \ref{fig:scale}.  The problem with a $1024 \times 1024$ mesh is simulated with $128$, $512$, $2048$, $8192$ threads. We observe a steady decrease of wall time with increased threads of GPU. The slope is  $-0.73$, suggesting a  good scalability of the algorithm.   We attribute the  deviation from the perfect slope of $-1$ to two reasons.  First, we use FFT which scales as  $O(n \log n)$ with system size, and second, operations such as sum and norm also takes communication and do not scale linearly with the number of nodes.   We note that the scaling improves for nonlinear problems as the local step 1 takes a larger fraction of the time.  The analogous result for liquid crystal elastomers is shown in Figure \ref{fig:lcess} of the Appendix and the slope is -0.80.
The scaling efficiency is  confirmed by weak scaling. The same configuration is studied with a $128 \times 128$ grid using $32$ threads, a $256 \times 256$ grid using $128$ threads, a $512 \times 512$ grid using $512$ threads, a $1024 \times 1024$ grid using $2048$ threads, and a $2048 \times 2048$ grid using $8192$ threads. 
Overall, the algorithm and GPU implementation show good parallel efficiency with system size.


\section{Microstructure evolution in liquid crystal elastomers} \label{sec:lce}

The second example concerns liquid crystal elastomers (LCEs), and demonstrates the ability of the proposed approach to address problems with evolving internal variables including those that are subject to constraints.  

\subsection{Liquid crystal elastomer formulation}

LCEs are synthetic materials made by incorporating rod-like nematic mesogens into the polymer chains of an elastomer \cite{warner2003liquid}. The combination of polymer elasticity and nematic ordering gives rise to exotic mechanical behavior.  At high temperatures, the mesogens are randomly oriented and the LCE is an isotropic rubbery solid.  However, on cooling, they undergo a phase transition where the steric interactions between the mesogens cause them to align in a particular direction.  The isotropic to nematic phase transition is also accompanied by a spontaneous elongation.  Thus, they have been proposed for applications as actuators  and for shape-morphing.  Further, if the LCE is synthesized in the isotropic phase, then the direction of nematic ordering can change freely, and this results in an unusual soft behavior.  We refer the reader to Warner and Terentjev \cite{warner2003liquid} for a comprehensive introduction.

We take the configuration in the isotropic state to be the reference configuration, but consider the material at a temperature below the phase transition temperature so that it is in the nematic phase.  The state of the material at a material point $x$ is then described by the deformation gradient $F(x)$, and a nematic director $n(x)$ that describes the orientation of the nematic mesogens in an infinitesimal volume around $x$.  
The material is typically incompressible and so $\det F =1$, and $n$ is a unit vector ($|n|=1$) since it describes an orientation.  

The free energy per unit volume of the material is given by,
\begin{equation} \label{eq:wlce}
W(F, n, \nabla n,x) =    W_{el}(F,n) + W_{ni}(F,n,x) + W_{F}(\nabla n)
\end{equation}
where the three terms describes three different physics. The first term,
\begin{align}
 W_{el}(F,n) = \frac{1}{2} \mu Tr(F^{\mathit{T}}  \ell^{-1}  F )  \quad  \text{where} \quad   \ell = r^{-1/3}(\delta - (r-1)n\otimes n)
\end{align}
describes the entropic elasticity of the polymer network  \cite{bladon1993transitions}.  $\mu$ is the shear modulus,  $\delta$ is identity tensor,  $r > 1$ represents the degree of nematic order that depends on temperature, and $\otimes$ represents the tensor product.  We take $r$ to be fixed since we fix temperature.  Note that if $r=1$, then $\ell = \delta$ and $W_{el}$ reduces to the neo-Hookean energy~\cite{fox}.  For $r>1$, the set of ground states ($W_{el} = 0$) corresponds to $F= R \ell_0^{1/2} Q, n= R e$ for rotations $R,Q$ and fixed unit vector $e$.  In other words, the material elongates along the director by factor $r^{1/3}$ and contracts perpendicular to it by factor $r^{-1/6}$, and the director is free to take any orientation.  The second term, 
\begin{align}
    W_{ni}(F,n,x) &=  \frac{1}{2} \mu \alpha Tr( (\delta - n_0(x) \otimes n_0(x) )  F^{\mathit{T}}  (n \otimes n)  F)
\end{align}
describes the `non-ideality' originating from non-uniformity in the cross-link density  \cite{biggins2008semisoft}.  $\alpha$ is the strength of the non-ideality and $n_0(x)$ is a fixed random unit vector field.  The non-uniformity in the cross-link density seeks to orient the director $n$ parallel to $F n_0$ at $x$, but this is a weak preference since $\alpha$ is typically small. Finally, the third term 
\begin{align}
    W_{F}(\nabla n) &=  \kappa |\nabla n|^2 =  \frac{1}{2} \kappa |\nabla (n \otimes n)|^2 
\end{align}
 is (a constant coefficient approximation of) Frank elasticity \cite{warner2003liquid, wojtowicz1975introduction}.  It reflects the preference of the directors to align spatially.   It is easy to verify that both forms of the expression shown are equivalent using the fact that $|n|=1$.  We note for later use that $\sqrt{\kappa/\mu}$ determines a length-scale of the domain wall and is typically $\mathcal{O}(10$nm) \cite{warner2003liquid}.  

The evolution is controlled by a dissipation potential which we take to be
\begin{equation}
D(\dot F, \dot n) = {1 \over 2} \nu_F |\dot F|^2 + {1 \over 2} \nu_n |\dot n|^2
\end{equation}
Note that this expression is not frame-indifferent  and there are a number of generalizations \cite{mos_siamjam_14}.  However, this is commonly used when the boundary conditions do not involve large rigid body rotations.  
The implicit time discretization of the evolution equation gives rise to the following variation problem (cf. \ref{eq:var})
\begin{equation} 
u^{k+1}, n^{k+1} = \argmin_{|n| = 1, \det \nabla u = 1}   \int_\Omega \left( W( \nabla u, n, \nabla n, x) + \Delta t 
D\left( {F - F^n \over \Delta t}, {n - n^k \over \Delta t} \right) \right) dx
\end{equation}
Since $F$ and $n$ satisfy constraints, one should consider the non-Euclidian metrics along the constraint manifold instead of the Euclidian metric in the embedding space.  However, the approximate expressions are accurate to first order. 
 
We discretize space using finite differences and solve the resulting equatons according to the massively parallel approach described in Sections \ref{sec:meth} and \ref{sec:imp} with two modifications.   First, the constraint of incompressibility, $\operatorname{det} F = 1$, is enforced in the local Step 1 using a Lagrange multiplier; while the constraint on the director, $|n|=1$ is enforced by introducing Euler angles.  Second, we also have a gradient of $(n \otimes n)$ in our functional.  We could proceed by introducing an auxiliary variable for $\nabla (n \otimes n)$ and using a constraint for it.  However, we have found that we obtain satisfactory results by treating this term explicitly.  All simulations are performed under periodic boundary conditions on the deformation gradient and the director. 

\subsection{Monodomain  LCE}

\begin{figure}
\centering
\includegraphics[width=\textwidth]{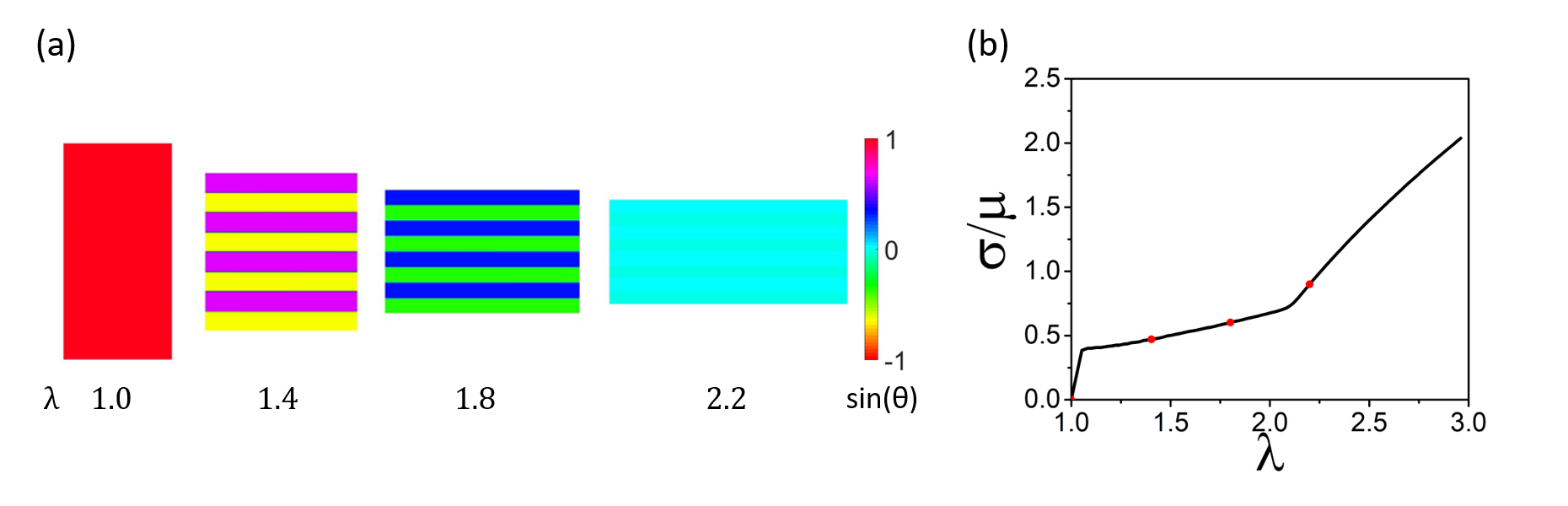}
\caption{Two-dimensional study of a monodomain LCE subjected to uniaxial stress.  (a) Evolution of the director ($\theta$ is the angle between the director and the horizontal loading direction). (b) Stress-stretch curve (stress is normalized by the modulus $\mu$). }
\label{fig:mono}
\end{figure}

We begin by studying a simple problem in two dimensions inspired by the experiments of K\"upfer and Finkelmann \cite{kf_dmc_91}.   We start with a monodomain specimen where $n_0 = e_2$ is uniformly in the vertical direction as shown in Figure \ref{fig:mono}.  We normalize the energy density with $\mu$ and take the rest of the parameters to be $r = 4$, $\alpha = 0.1$, $\kappa / \mu = 7.63 \times 10^{-6} $ ($\kappa$ in non-dimensional length units).  We set dissipation to zero taking $\nu_n=\nu_F=0$ so that we solve for equilibrium at each time step.  Finally, we subject the specimen to an average deformation gradient $\langle F_{11} \rangle = \lambda, \langle F_{12} \rangle = \langle F_{21} \rangle = 0$ and $\langle F_{22} \rangle$ free and solve it with a resolution of $256 \times 256$.  We use the previous configuration along with a small periodic perturbation of the order $10^{-4}$ in $F$ as the initial guess at each time step.  The resulting domain patterns are shown in Figure \ref{fig:mono}(a) while the stress-stretch curve is shown in Figure \ref{fig:mono}(b).   We observe the semi-soft behavior and stripe domains consistent with the experimental observations.  The director is initially aligned with $n_0$ and there is no stress.  At small applied stretch, the LCE reacts elastically as the non-ideal term keeps the director $n$ aligned with $n_0$.  At a critical stretch, the director can align, but doing so leads to a shear inconsistent with the imposed condition on the deformation gradient.  Therefore, it forms stripe domains where the director rotates in opposite directions in alternating stripes; the two regions have the same stretch but opposing shear so that they can satisfy the imposed average deformation gradient condition.  The spacing depends on the perturbation and $\kappa$.  The formation of stripe domains is accompanied by a softening in the stress-stretch curve.  The director continues to rotate as the stretching continues until it is fully rotated to the horizontal when both domains merge (since the sign of the director has no meaning).  The stress-stretch curve then stiffens as the material responds elastically.  All of this is consistent with the observations of K\"upfer and Finkelmann \cite{kf_dmc_91} and prior theoretical considerations \cite{warner2003liquid}.

\subsection{Polydomain LCE}

In this section, the parameters are $r = 7.71$, $\mu =  23.63$ kPa, $\alpha$ = 0.06, $\kappa =  3.61 \times 10^{-9} $N , 
$\nu_F = 2.65$ kPa.s and $\nu_n=0.005$ kPa.s unless otherwise specified.  We conduct our simulations on a (1 $\mu$m)$^3$ cubic unit cell with a 128$^3$ resolution at a strain rate of 1 s$^{-1}$ with time steps of $0.02$s.

\paragraph{Polydomain material}

\begin{figure}
\centering
\includegraphics[width=0.9\textwidth]{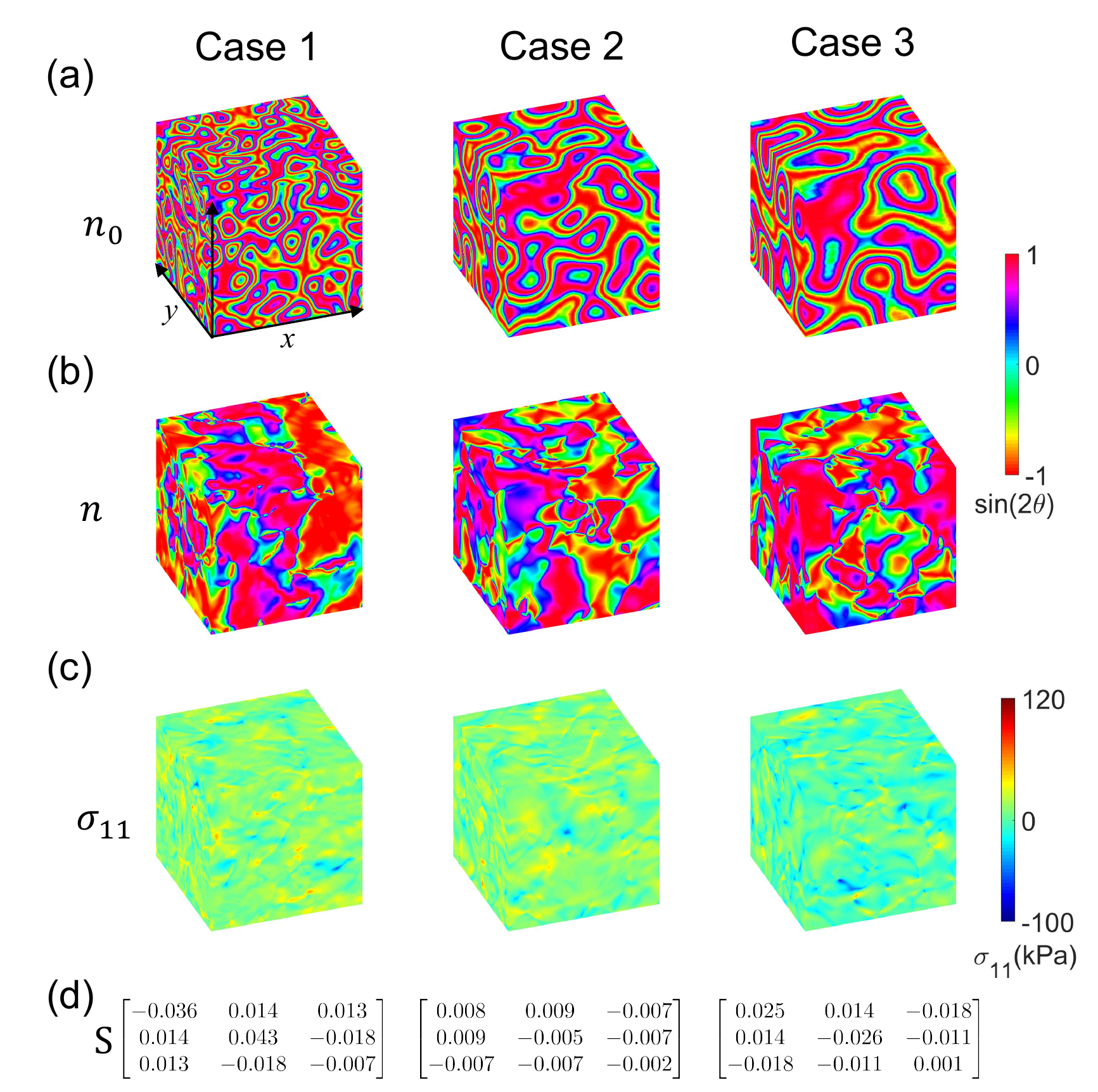}
\caption{Creating the initial polydomain material.  (a) Distribution of the preferred director $n_0$ with fluctuations on three length-scales, (b) Distribution of director $n$ after relaxation, (c) Internal stress distribution and (d) Orientation tensor (\ref{eq:orient}) after relaxation ($\theta$ is the in-plane angle of mesogens w.r.t. $x-$axis.) }
\label{fig:init}
\end{figure}

We begin by exploring the effect of the random director field $n_0$ on the initial configuration.  Figure \ref{fig:init}(a) shows three `random' director fields $n_0$ with fluctuations on a different length-scale.  All three of them are generated by starting with the same Gaussian random field of angles and then filtering to different length scales. With an initial guess of $n=n_0, F=I$, we let the system relax under zero average stress and we obtain the director field $n$ shown in Figure \ref{fig:init}(b).  We see that the relaxed $n$ does not follow $n_0$, and the system is internally stressed as shown in Figure \ref{fig:init}(c).   Interestingly, the length-scale on which $n$ fluctuates is similar in each of the three cases.  It is larger than the length-scale of $n_0$ fluctuation in the first two cases, but similar in the third.  Importantly, $n$ remains largely equi-distributed.  It is useful to look at the orientation tensor 
\begin{equation} \label{eq:orient}
S = {3 \over 2} \left( {1 \over 3} I - \langle n \otimes n \rangle \right)
\end{equation}
where $ \langle \cdot \rangle$ denotes the average over the computational domain \cite{warner2003liquid}.  An equi-distributed director field leads to $S = 0$.  Figure \ref{fig:init}(d) shows that $n$ is essentially equi-distributed. 

We may understand this initial relaxation as follows.  Recall that the non-ideal term prefers that the director $n$ follow the prescribed $n_0$ and the elastic energy  prefers an elongation along director.  However, this resulting distortion field may not be compatible leading to elastic energy.  Further, the Frank energy penalizes the fluctuations in the director field.  Thus, the competition between these three terms drives the relaxation, and the resulting director pattern is a compromise between them.  

\begin{figure}
\centering
\includegraphics[width=0.9\textwidth]{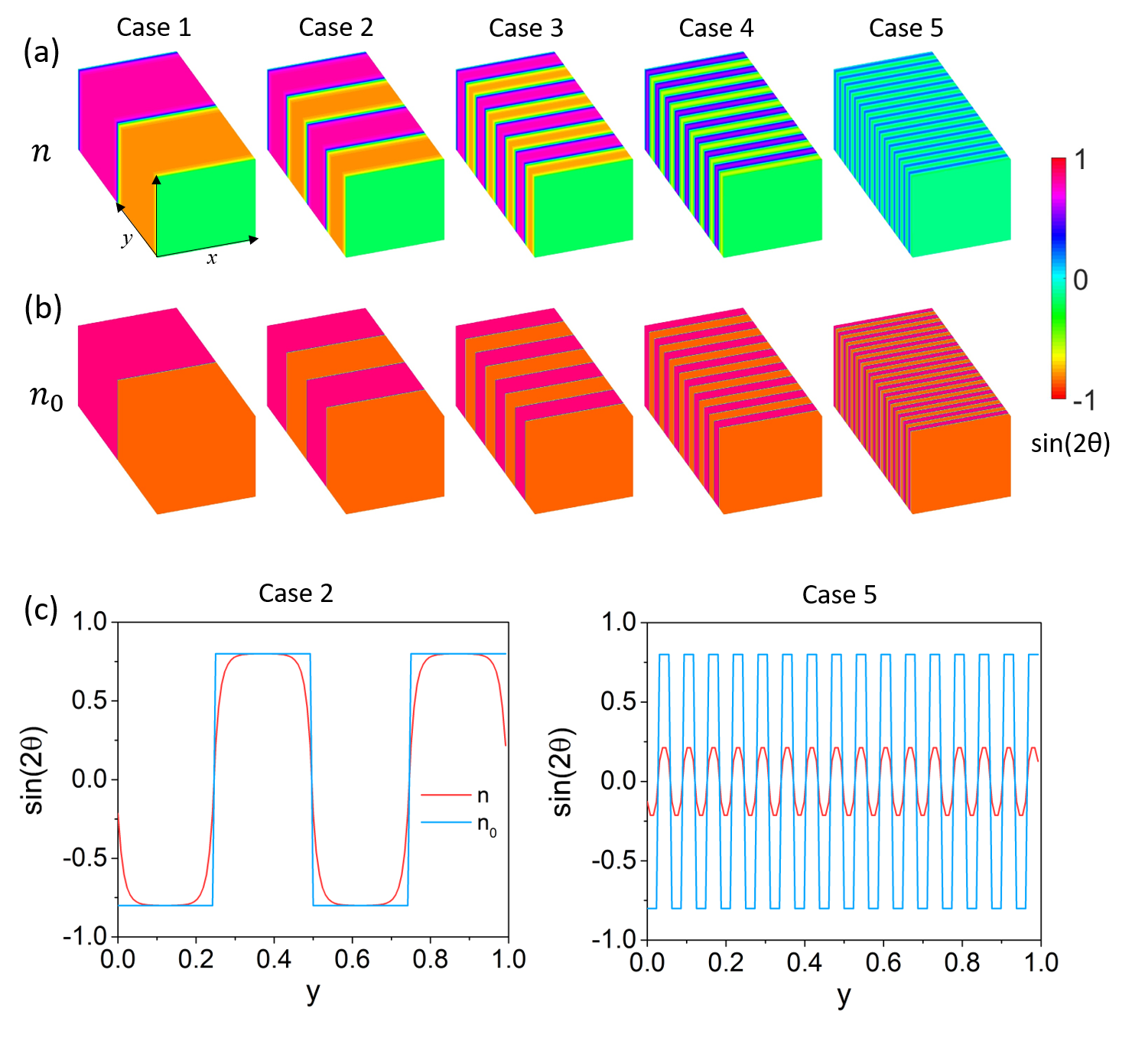}
\caption{Relaxation of a stripe polydomain.  (a) Distribution of the compatible preferred director $n_0$ with multiple length-scales, (b) Distribution of director $n$ after relaxation ($\theta$ is the in-plane angle of mesogens w.r.t. $x-$axis and $\sin 2 \theta$ is the product of the $x$ and $y$ components of the director.).   (c,d) Details of the director field: the $y$-component of the director versus the $y$-coordinate for the second (c) and last cases (d) of (a,b).}
   	\label{fig:stripe}
\end{figure}

To verify this, we start with a `compatible' initial director field $n_0$ that takes two distinct values $n_0^{\pm} = \{\pm 0.42, 0.91, 0\}$ in alternating stripes as shown in Figure \ref{fig:stripe}(a,c).  It is easy to verify that the two two corresponding spontaneous stretches $ (\ell(n_0^\pm)^{1/2}$ are kinematically compatible across an interface with normal $e_2$, i.e., we can find a rotation $Q$ and a vector $a$ such that $Q (\ell(n_0^+)^{1/2} - (\ell(n_0^-)^{1/2} = a \otimes e_2$.  Thus, the director field $n=n_0$ and deformation with gradient $F= (\ell(n))^{1/2}$ is admissible and minimizes the sum of the first two terms in the energy (\ref{eq:wlce}).   Thus we expect the solution to follow $n=n_0,  F= (\ell(n))^{1/2}$ except close to the interface where we expect a transition layer with thickness of the order $\sqrt{\kappa/\mu}$.    We study five cases with distinct length-scales.  As before, we start with $n=n_0, F=I$, and let the system relax under zero average stress.  We obtain the director field $n$ shown in Figure \ref{fig:stripe}(b,c). If the length-scale is is sufficiently large (the first four cases), then $n$ follows $n_0$ except near the interface where we see a transition layer as we expect.  At smaller length-scales the Frank elasticity prevents $n$ from completely relaxing to $n_0$; in other words the interfaces dominate.  This calculation shows that kinematic compatibility drives the relaxation with the Frank elasticity setting the length-scale.

\paragraph{Uniaxial and biaxial deformation}

\begin{figure}
\centering
\includegraphics[width=0.9\textwidth]{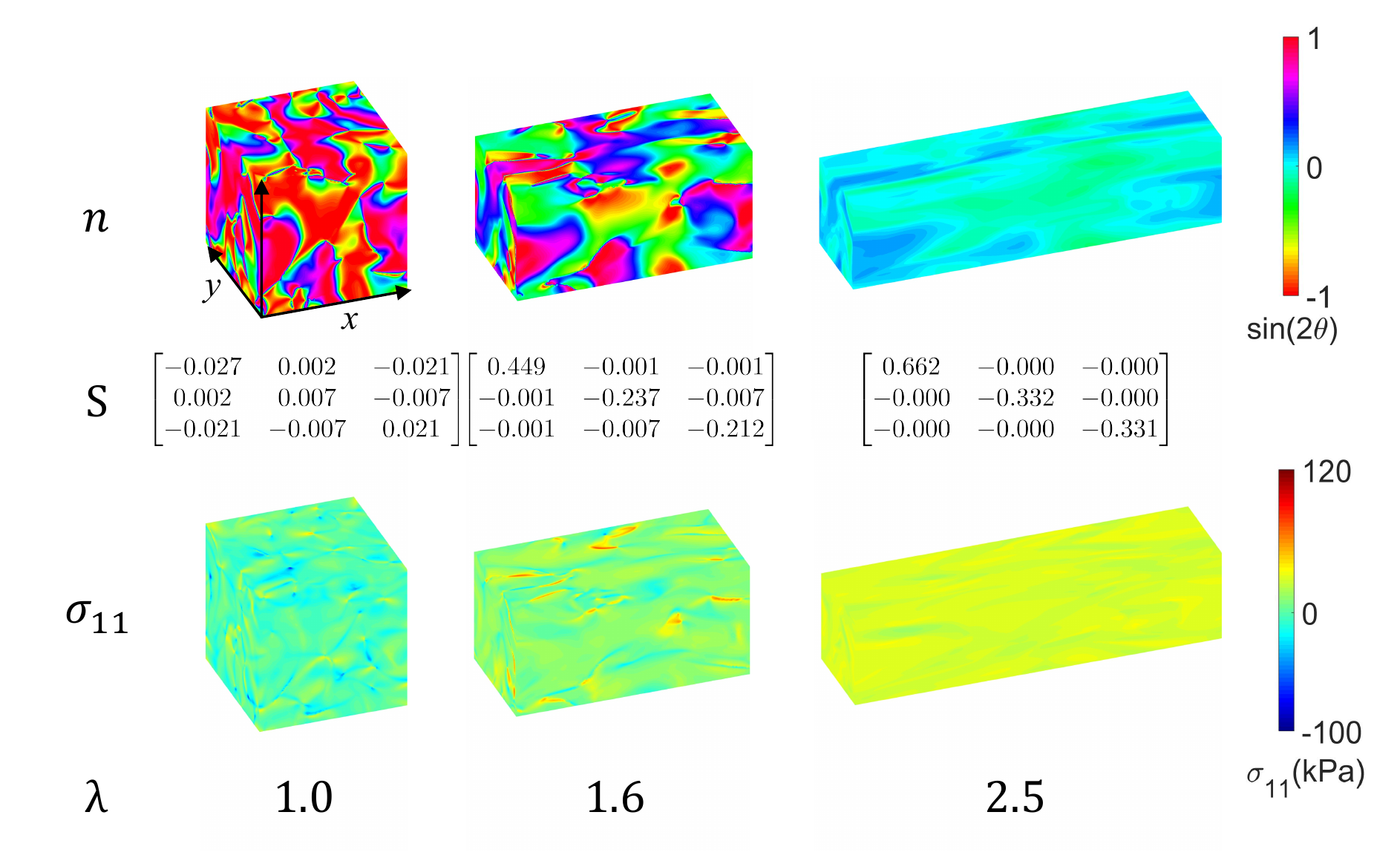}
\caption{Uniaxial stress (UNI).  Evolution of the director, the orientation tensor (\ref{eq:orient}) and the stress with stretch.}
\label{fig:uni}
\end{figure}

\begin{figure}
\centering
\includegraphics[width=0.9\textwidth]{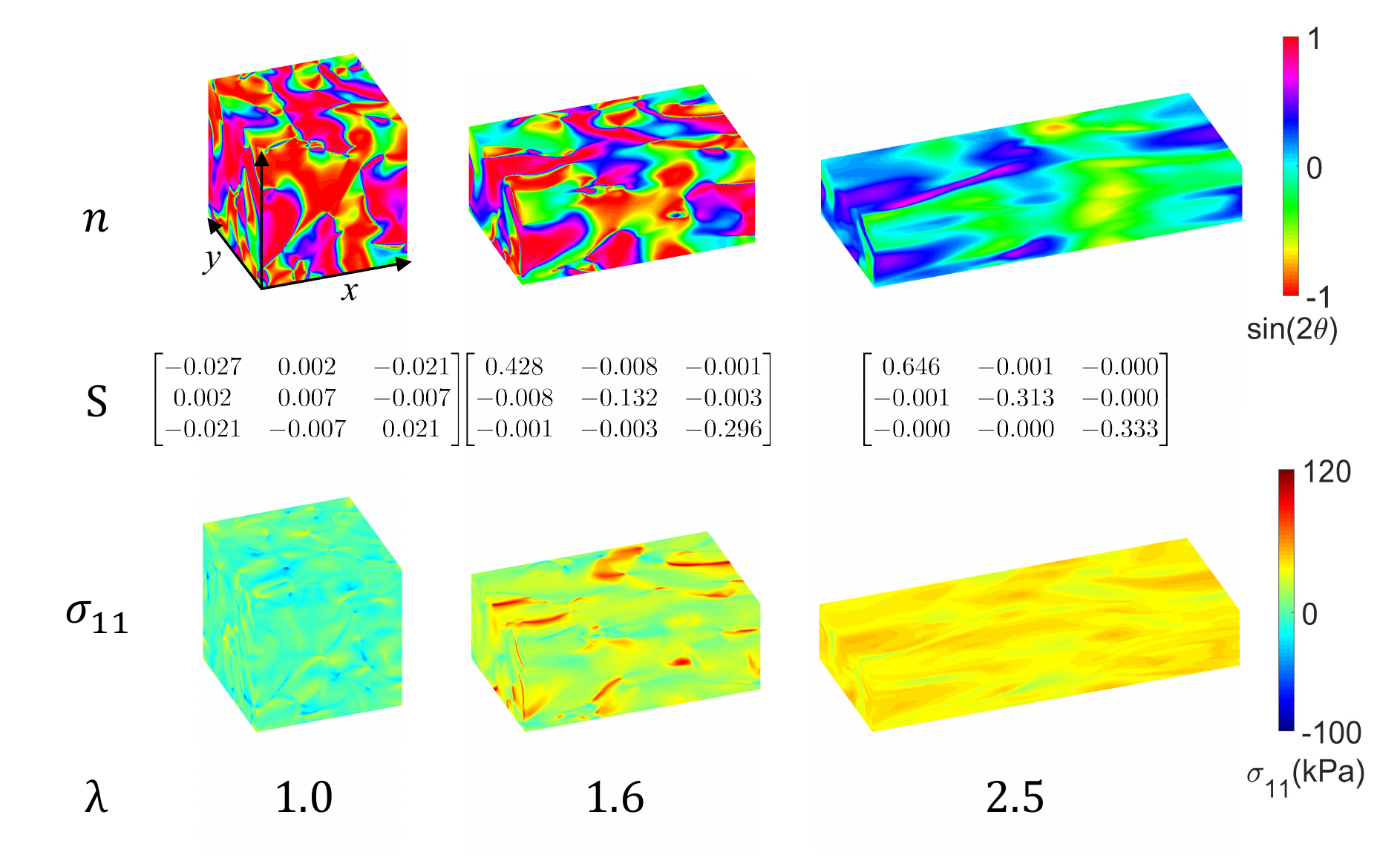}
\caption{Uniaxial stretch of a sheet in plane stress (PE).  Evolution of the director, the orientation tensor (\ref{eq:orient}) and the stress with stretch.}
\label{fig:pe}
\end{figure}

\begin{figure}
\centering
\includegraphics[width=0.9\textwidth]{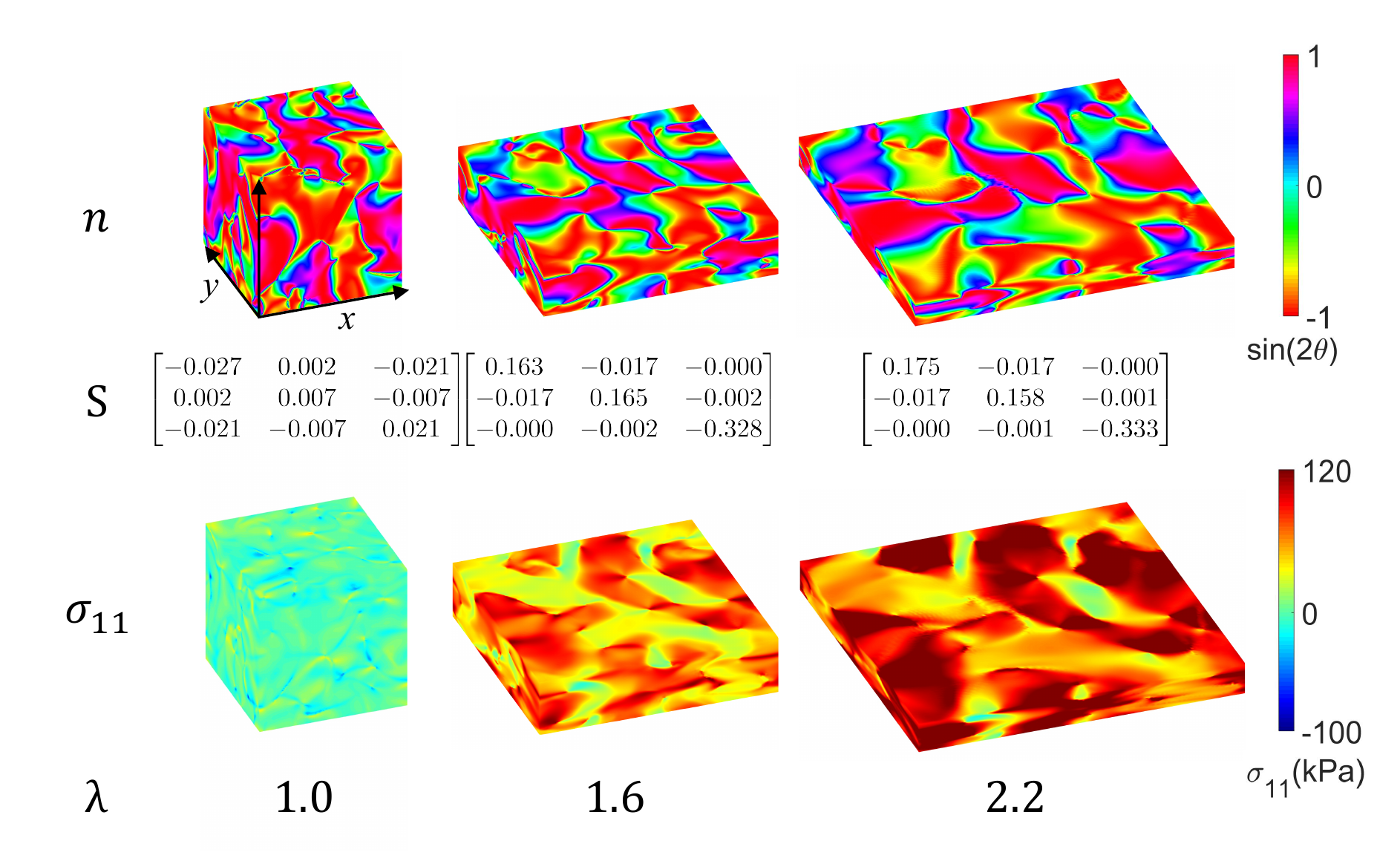}
\caption{Equi-biaxial stretch of a sheet in plane stress (EB).  Evolution of the director, the orientation tensor (\ref{eq:orient}) and the stress with stretch.}
\label{fig:eb}
\end{figure}

\begin{figure}
\centering
\includegraphics[width=0.95\textwidth]{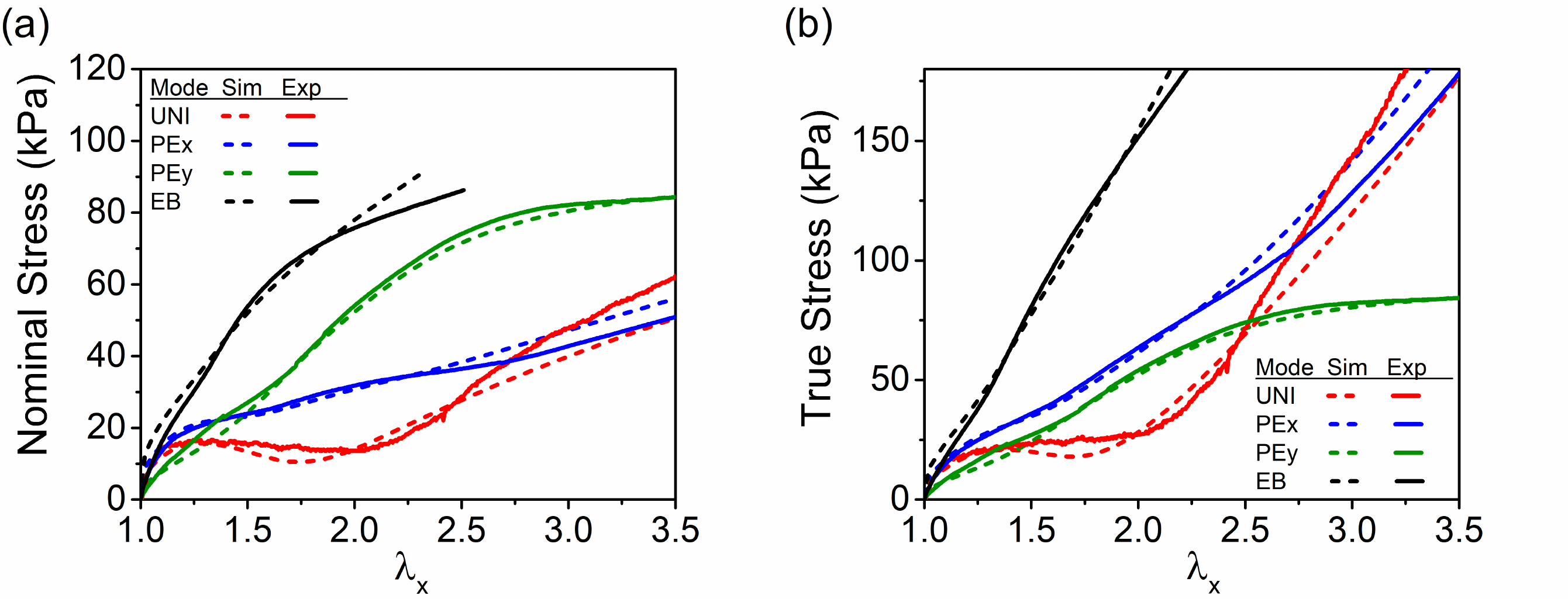}
\caption{Stress vs. stretch for various loading protocols (UNI, PE and EB) obtained by both simulation (dashed line) and experiment (solid line).  (a) Nominal stress vs. stretch and (b) True stress vs. stretch. }
   	\label{fig:ss}
\end{figure}

These simulations are motivated by the experiments reported in Tokumoto {\it et al.} \cite{tetal_20}.  They took $65 \times 65 \times 0.7$ mm sheets and subjected them to uniaxial stress and biaxial stretch protocols.  To replicate these experiments, we start with a relaxed polydomain specimen prepared as described above and subject it to three loading protocols.
\begin{itemize}
\item UNI: uniaxial stress ($\langle F_{11} \rangle$ is prescribed while all other components are free);
\item PE: uniaxial stretch of a sheet in plane stress ($\langle F_{11} \rangle= \lambda_x, \langle F_{12} \rangle=0, \langle F_{22} \rangle = 1$ while the other components are free);
\item EB: equi-biaxial extension of a sheet in plane stress ($\langle F_{11} \rangle = \langle F_{22} \rangle =\lambda_x, \langle F_{12} \rangle = 0$ while the other components are free).
\end{itemize}
Note that PE and EB are mixed boundary conditions on the unit cell.  Since $\langle F_{i3} \rangle$ and $\langle F_{3i} \rangle$ are left free and part of the minimization, the average tractions on the faces with normal $e_3$ are zero, and the average tractions on the other faces are planar.  This corresponds to plane stress.   At the same time, the average planar stretch is prescribed.  The evolution of the director and the stress are shown in Figures \ref{fig:uni}, \ref{fig:pe} and \ref{fig:eb}.   The macroscopic stress-strain curve is shown in Figure \ref{fig:ss} and compared to experimental observations.

We observe that the director pattern, residual stress and overall stress-strain curve are very different in the different loading scenarios.  In uniaxial stress (UNI, Figure \ref{fig:uni}), the directors rotate until they all eventually align.  This is similar to the situation in the ideal material (Figure \ref{fig:mono}) though the actual patterns are more complicated.  Any residual fluctuations are small and this is also reflected in the small stress heterogeneity.  This microstructure evolution leads to a soft plateau in the macroscopic stress-strain response which eventually stiffens when all the directors are aligned.

In the case of uniaxial stretch in plane stress (PE, Figure \ref{fig:pe}), the directors again try to rotate to the direction of elongation, but are prevented from doing so by the lateral constraint.  Therefore, significant amounts of residual microstructure and some residual stress persist.  Further, the macroscopic stress-strain response shows only a small plateau.  The macroscopic stress-strain response also shows another rather interesting feature.  The nominal stress and thus the applied force in the stretching ($x$) direction is smaller than those in the constrained ($y$) direction.  This is counter-intuitive, and different from the behavior of ordinary elastomers.   The reason for it is evident by examining the true or Cauchy stress: we observe the true stresses are (almost) equal in the two directions despite the fact that the stretches in the two directions.  In other words, we are in a state of equi-biaxial stress with no shear stress.  Cesana {\it et al.} \cite{cpb_arma_15} predicted a region of equi-biaxial stress in ideal materials ($\alpha = \kappa =0$).  This behavior remains in non-ideal materials.

In the case of equi-biaxial stretch (EB, Figure \ref{fig:eb}), the directors orient gradually to become planar, but there is little, if any, evolution beyond that.  There is also consequently significant heterogeneity in the state of stress.  Together, the three results show that shear of unequal stretch drives microstructure evolution.  

Finally, the results of the simulations above are in agreement with the observations of Tokumoto {\it et al.} \cite{tetal_20}.  This is shown by comparing the stress-stretch relations in Figure \ref{fig:ss}.  The agreement is excellent and this is all the more remarkable because all simulations are carried out with only six parameters.

\section{Conclusion} \label{sec:conc}
 
We have presented an approach to solving problems of computational micromechanics that is amenable to massively parallel calculations through the use of graphical processing units and other accelerators.  The approach is based on splitting the solution operator in a manner that exploits the structure of continuum models that combine linear and universal physical laws (kinematic compatibility, balance laws), and nonlinear but local constitutive relations. We demonstrate it with two examples. The first concerns the long wavelength instability of finite elasticity, and allows us to verify the approach against previous numerical simulations.  We also use this example to study convergence and parallel performance.  The second example concerns microstructure evolution in liquid crystal elastomers and provides new insights into some counter-intuitive properties of these materials.  We use this example to validate the model and the approach against experimental observations.  We note that the method can be applied to a variety of problems.  These include crystal plasticity, martensitic phase transformations, twinning, precipitation and Landau-Ginzburg models since these problems lead to equations of the form (\ref{eq:elas},\ref{eq:int}).

We conclude with a discussion of extensions and open issues.  The implementation and examples presented here concern periodic boundary conditions which enabled the use of fast Fourier transforms to solve Poisson's equation. However, periodic boundary conditions are not inherent to this approach.  The key issue is the solution of Poisson's equation, and there are a number of parallel iterative approaches that have been implemented with accelerators \cite{retal_apc_14}.   The problem of liquid crystal elastomers showed that we can incorporate point-wise constraints (incompressibility and prescribed norm on a vector internal variable) naturally in this method.  It is possible to extend this approach to problems like fracture and contact where one has inequality constraints.  Finally, one can extend this method to phenomena that include higher derivatives by introducing additional auxilliary variables.

An important open question is the convergence of the algorithm and error estimates.   We have noted in Section \ref{sec:form} that there are partial results in the case of convergence.  However, the general case where $W$ is quasi-convex in $F$ and convex in the internal variables remains open.  Further, systematic analysis of the error remains a topic for the future.

\newpage
\section*{Appendix}

\paragraph{Equilibrium condition}
We show that the dual feasibility ensures satisfaction of the equilibrium equation of mechanics.  We begin with the case when the local problem (step 1) is solved exactly. 
Consider a smooth test function $\varphi: \Omega \to {\mathbb R}^3$ that vanishes on the boundary.  Multiply (\ref{eq:helm}) with $\varphi$, integrate over the domain and use the divergence theorem to obtain
\begin{equation}
\int_\Omega \nabla \varphi \cdot \left( \rho(F^{n+1} - \nabla u^{n+1}) - \Lambda^n \right) dx = 0.
\label{eqn:compat}
\end{equation}
Now multiply (\ref{eq:eq}) with $\nabla \varphi$ and integrate over the domain to obtain
\begin{equation}
\int_\Omega \nabla \varphi \cdot \left(W_F(F^{n+1}, \lambda^{n+1}, x) - \Lambda^n + \rho( \nabla u^n - F^{n+1}) \right) dx = 0.
\end{equation}
Subtract one from the other and we obtain
\begin{equation}
\int_\Omega \nabla \varphi \cdot W_F(F^{n+1}, \lambda^{n+1}, x) dx = \rho \int_\Omega \nabla \varphi \cdot (\nabla u^{n+1} - \nabla u^n) dx.
\end{equation}
By the dual feasibility (\ref{eq:check})$_2$ and the Cauchy-Schwarz inequality, the right hand side above goes to zero.  Further, the left hand side converges to the weak from of the equilibrium equation since this holds for arbitrary $\varphi$.

When the local step is not exact, we rewrite (\ref{eqn:compat})
\begin{align} \label{eq:inexact}
&\int_\Omega \nabla \varphi \cdot W_F(F^{n+1}, \lambda^{n+1}, x) dx =  \nonumber \\
&\rho \int_\Omega \nabla \varphi \cdot (\nabla u^{n+1} - \nabla u^n) dx + \int_\Omega \nabla \varphi \cdot \left(W_F(F^{n+1}, \lambda^{n+1}, x) - \Lambda^n + \rho( \nabla u^n - F^{n+1}) \right) dx.
\end{align}
The first term on the right is bounded by dual feasibility (\ref{eq:check})$_2$ and the Cauchy-Schwarz inequality as before, and the second term is bounded by the local error estimate and the Cauchy-Schwarz inequality.  Thus the weak form of the equilibrium equation holds.

\begin{figure}[t!]
\centering
\includegraphics[width=0.9\textwidth]{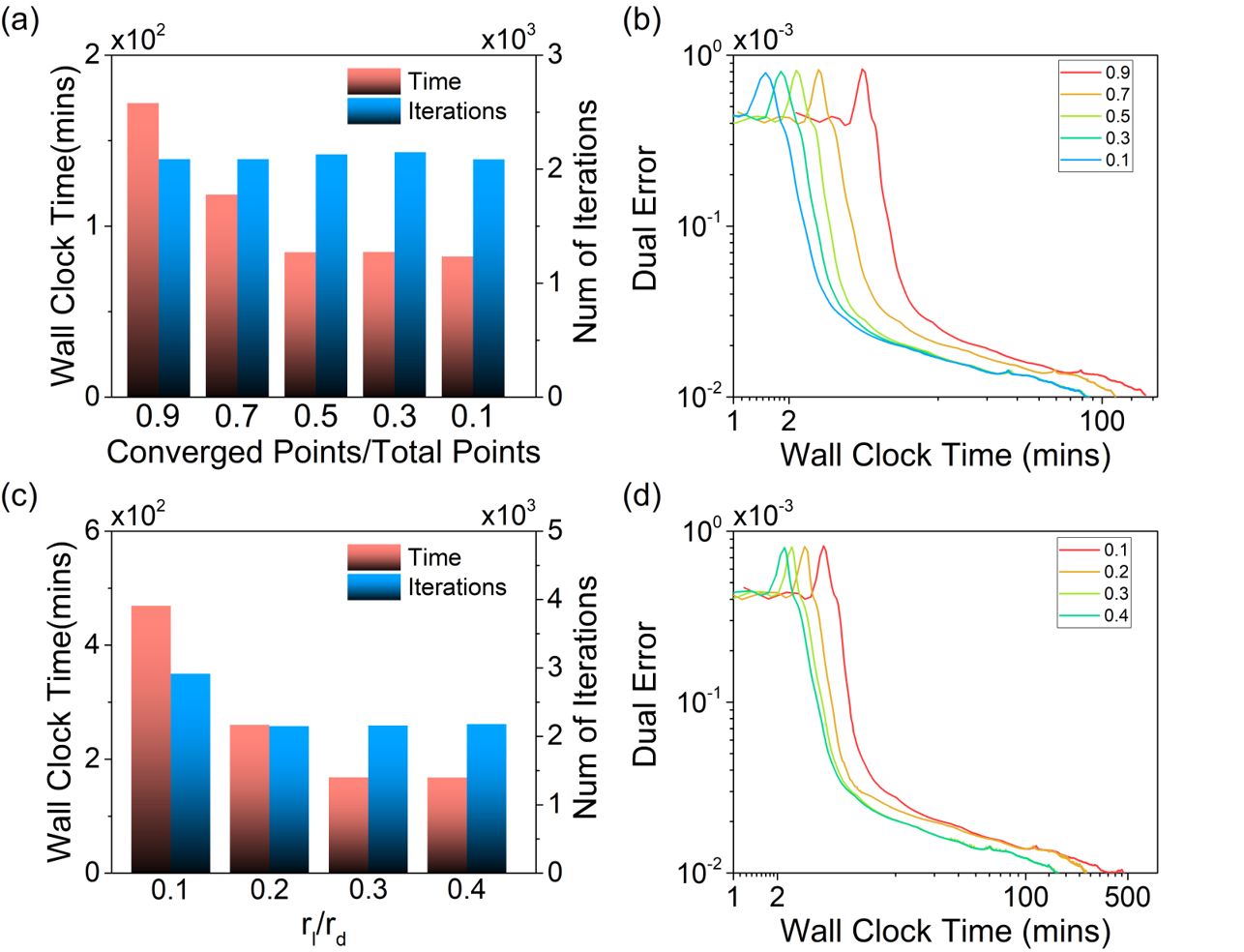}
\caption{Performance with an approimate solution of the local problem in the case of liquid elastomers.  (a,b) Local convergence on a fixed fraction of spatial points: (a) Wall clock time and number of global iterations for global convergence for various fractions.  (b)  The global dual error versus wall clock time for various fractions.  (c,d) Fixed ratio of local ($r_l$) to global dual ($r_d$) residual: (c) Wall clock time and number of global iterations for global convergence for various ratios and (d) The global dual error versus wall clock time for various ratios.}
\label{fig:lceinexact}
\end{figure}

\paragraph{Convergence and performance in the case of LCEs}
Figure \ref{fig:lceinexact} shows the performance of the algoirthm when we have an approximate solution of the local problem (step 1).  As in the case of the elasticity problem discussed in Section \ref{sec:num}, we find that the approximate solution provides savings in time without affecting the overall global convergence.

\begin{figure}[b!]
\centering
\includegraphics[width=0.4\textwidth]{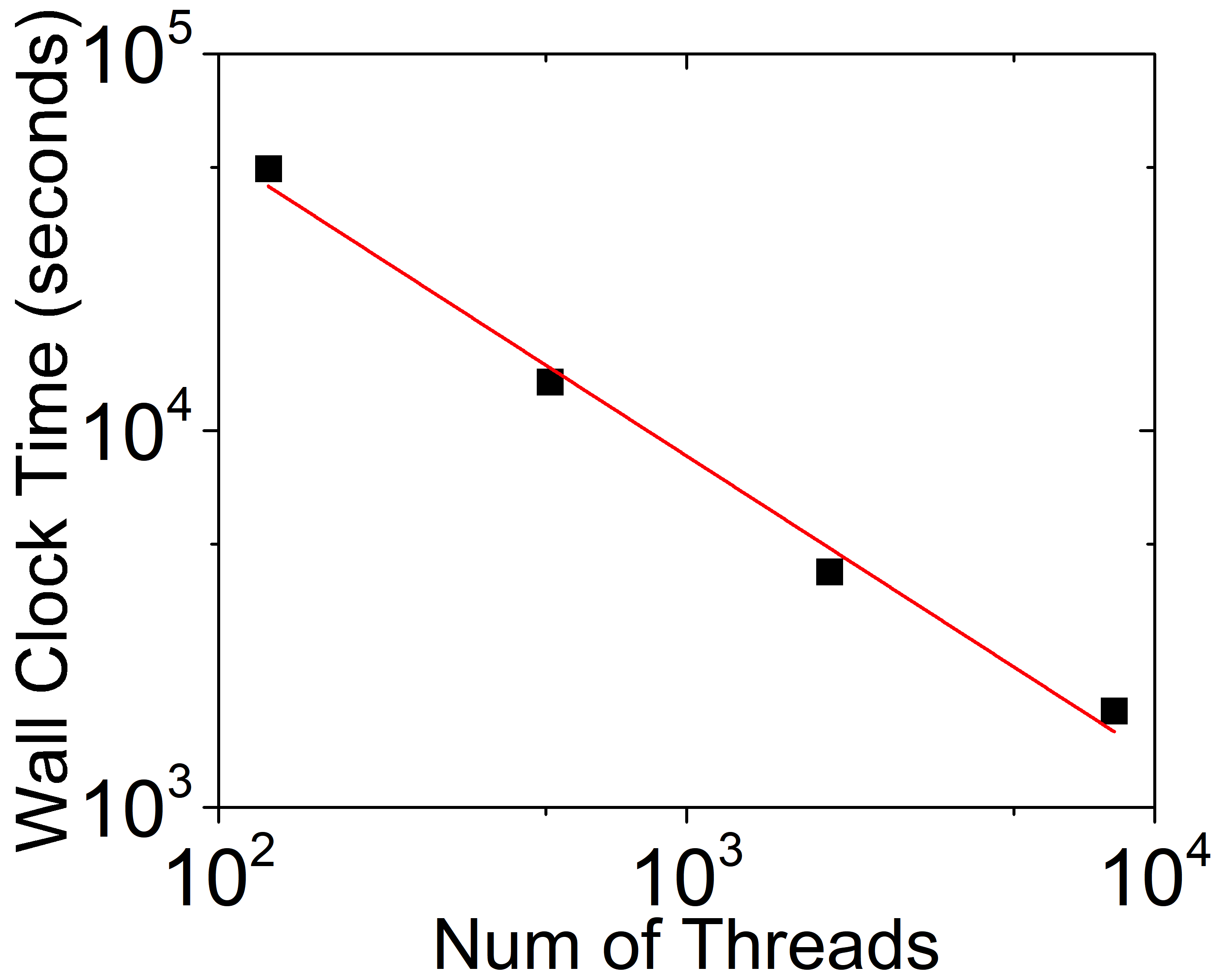}
\caption{Strong scaling in the case of liquid crystal elastomers.}
\label{fig:lcess}
\end{figure}

Figure \ref{fig:lcess} shows the strong scaling in the case of liquid crystal elastomers.  The slope is -0.80 which is very good, and in fact better than that observed in the case of elasticity.  This is because the local step 1 which scales linearly takes a larger fraction of time compared to the case of elasticity.  We have not performed the analysis of weak scaling since the specification of $n_0$ typically depends on the spatial resolution and therefore one-to-one comparison between simulations with different resolutions is not possible.

\section*{Acknowledgement}
We are delighted to acknowledge many stimulating discussions with Pierre Suquet (concerning FFT algorithms) and Kenji Urayama (concerning LCEs).  The latter also generously provided us with experimental data shown in Figure \ref{fig:ss}.  We gratefully acknowledge the support of the US Air Force Office for Scientific Research through the MURI grant number MURI grant FA9550-16-1-0566.  The computations presented here were performed at the High Performance Computing Center of California Institute of Technology.

\end{document}